\documentclass[aps,prl,showpacs,twocolumn,preprintnumbers,superscriptaddress,nofootinbib,floatfix,10pt]{revtex4-1}
\usepackage{amsfonts,amssymb,stmaryrd,latexsym,amsmath,braket}
\usepackage{graphicx,subfigure}
\usepackage{comment}
\usepackage{times}
\usepackage{slashed}
\usepackage{bm}
\usepackage{braket}
\usepackage{chapterbib}
\usepackage{color}
\usepackage{longtable}
\usepackage{hyperref}
\hypersetup{
    colorlinks=true,
    linkcolor=blue,
    citecolor=blue,
    urlcolor=blue
}
\usepackage{bibunits}

\newcommand{\beginsupplement}{%
        \setcounter{table}{0}
        \renewcommand{\thetable}{S\arabic{table}}%
        \setcounter{figure}{0}
        \renewcommand{\thefigure}{S\arabic{figure}}%
        \setcounter{equation}{0}
        \renewcommand{\theequation}{S\arabic{equation}}%
     }

\makeatletter
\let\saved@includegraphics\includegraphics
\AtBeginDocument{\let\includegraphics\saved@includegraphics}
\makeatother

\graphicspath{{figures/}}
\begin{document}

\begin{bibunit}[apsrev4-2]

\title{{\em Ab initio} study of the beryllium isotopes $^{7}$Be to $^{12}$Be}

\author{Shihang Shen}
\affiliation{Peng Huanwu Collaborative Center for Research and Education, International Institute for Interdisciplinary and Frontiers, Beihang University, Beijing 100191, China}
\affiliation{School of Physics, Beihang University, Beijing 102206, China}
\affiliation{Institute for Advanced Simulation (IAS-4), Forschungszentrum J\"{u}lich, D-52425 J\"{u}lich, Germany}

\author{Serdar Elhatisari}
\affiliation{Interdisciplinary Research Center for Industrial Nuclear Energy (IRC-INE), King Fahd University of Petroleum and Minerals (KFUPM), 31261 Dhahran, Saudi Arabia}
\affiliation{Faculty of Natural Sciences and Engineering, Gaziantep Islam Science and Technology University, Gaziantep 27010, Turkey}
\affiliation{Helmholtz-Institut f\"{u}r Strahlen- und Kernphysik and Bethe Center for Theoretical Physics, Universit\"{a}t Bonn, D-53115 Bonn, Germany}

\author{Dean Lee}
\affiliation{Facility for Rare Isotope Beams and Department of Physics and Astronomy, Michigan State University, East Lansing, MI 48824, USA}

\author{Ulf-G.~Mei{\ss}ner}
\email{meissner@hiskp.uni-bonn.de}
\affiliation{Helmholtz-Institut f\"{u}r Strahlen- und Kernphysik and Bethe Center for Theoretical Physics, Universit\"{a}t Bonn, D-53115 Bonn, Germany}
\affiliation{Institute for Advanced Simulation (IAS-4), Forschungszentrum J\"{u}lich, D-52425 J\"{u}lich, Germany}
\affiliation{Tbilisi State University, 0186 Tbilisi, Georgia}
\affiliation{Peng Huanwu Collaborative Center for Research and Education, International Institute for Interdisciplinary and Frontiers, Beihang University, Beijing 100191, China}

\author{Zhengxue Ren}
\affiliation{Institute for Advanced Simulation (IAS-4), Forschungszentrum J\"{u}lich, D-52425 J\"{u}lich, Germany}

\begin{abstract}
We present a systematic \textit{ab initio} study of the low-lying states in beryllium isotopes from \( ^7\text{Be} \) to \( ^{12}\text{Be} \)
using nuclear lattice effective field theory with the N\(^3\)LO interaction. Our calculations achieve good agreement with
experimental data for energies, radii, and electromagnetic properties. We introduce a novel, model-independent method to quantify nuclear shapes,
uncovering a distinct pattern in the interplay between positive and negative parity states across the isotopic chain.
By combining Monte Carlo sampling of the many-body density operator
with a novel nucleon-grouping algorithm, the prominent two-center cluster structures, the emergence of one-neutron halo,
complex nuclear molecular dynamics such as $\pi$-orbital and $\sigma$-orbital, emerge naturally.
\end{abstract}
\maketitle
\date{today}

\section{Introduction}

Beryllium isotopes are pivotal in nuclear structure studies due to their diverse phenomena, including clustering, halo structures, and
the breakdown of conventional shell closures. For instance, \( ^7\text{Be} \) plays a significant role in Big Bang nucleosynthesis and
nuclear astrophysics by influencing the primordial abundances of light elements \cite{Adelberger:1998qm,Adelberger:2010qa}. The
unbound \( ^8\text{Be} \), which decays into two alpha particles with a long lifetime, exemplifies nuclear instability and clustering effects.
Moving along the isotopic chain, \( ^9\text{Be} \) and \( ^{10}\text{Be} \) are renowned for their pronounced molecular-like structures \cite{Li:2023msp}. The
neutron-rich \( ^{11}\text{Be} \) is particularly notable for its ground-state parity inversion and halo structure, challenging traditional
shell-model predictions and providing insights into weakly bound systems \cite{Calci:2016dfb,Talmi:1960zz}. Similarly, \( ^{12}\text{Be} \)
exhibits the disappearance of the \( N = 8 \) shell closure, highlighting the role of intruder configurations and shape coexistence
\cite{Morse:2018ojw,Chen:2018hpx,Pain:2005xw,Navin:2000zz}.
These rich and varied phenomena underscore the need for comprehensive theoretical
frameworks capable of capturing the complex interplay of clustering, shell evolution, and continuum effects in beryllium isotopes.

A variety of theoretical approaches have been employed to study these isotopes, with significant emphasis on cluster structures, such as
Antisymmetrized Molecular Dynamics (AMD)~\cite{Kanada-Enyo:2014sqb,Kanada-Enyo:2002owv}, Fermionic Molecular Dynamics (FMD)~\cite{Krieger:2012jx},
molecular-orbital models~\cite{Itagaki:1999vm,von1996two}, the Tohsaki–Horiuchi–Schuck–Röpke (THSR) wave function
approach~\cite{Lyu:2021ykp,Lyu:2015ika,Lyu:2014ewa},
and other cluster models~\cite{Zhao:2022oik, Fukui:2020ylj, DellaRocca:2018mrt, Romero-Redondo:2008hux, Descouvemont:2002mnw}.
These methods have effectively captured cluster structures, molecular configurations, and the influence of valence neutrons in beryllium isotopes.
See also the recent reviews \cite{Kimura:2016fce, Funaki:2015uya, von2006nuclear}.
In parallel, density functional theory has been employed to predict the formation of alpha clusters bonded by excess neutrons,
highlighting the significant role of clustering in these systems \cite{Geng:2024oex,Geng:2023pao,Ebran:2014pda}.
\textit{Ab initio} methods such as the Gamow shell model \cite{LinaresFernandez:2023bzu}, Green's function Monte Carlo
\cite{Carlson:2014vla,McCutchan:2009th,Wiringa:2000gb}, the resonating group method
\cite{Kravvaris:2017nyj}, Monte Carlo shell model (MCSM) \cite{Yoshida:2013dwa,Liu:2011xv}, and no-core shell
model (NCSM) \cite{McCoy:2024kah,Caprio:2021azr,Vorabbi:2019imi,Calci:2016dfb,Forssen:2004dk} have been instrumental in providing
a microscopic understanding of nuclear structure, clustering, and reaction dynamics in beryllium isotopes.

Comprehensive reviews have highlighted the importance of clustering phenomena in light nuclei and their impact on nuclear structure,
reaction dynamics, and astrophysics \cite{Lombardo:2023eht,gnoffo2022clustering,Kanada-Enyo:2018fjk,Fortune:2018ees,Freer:2017gip}.
These works emphasize the coexistence of cluster and shell-model features in neutron-rich isotopes and discuss the challenges in
fully understanding the underlying mechanisms.
While these theoretical methods have significantly advanced our understanding, a systematic \textit{ab initio} study encompassing energies,
radii, electromagnetic properties, and geometric structures across the beryllium isotopes is still lacking.
Moreover, the identification of cluster structure and molecular orbitals from the
full $A$-body wave function remains a question.

Recent advancements in nuclear
lattice effective field theory (NLEFT) offer promising avenues for such comprehensive investigations. The introduction of wavefunction
matching techniques within NLEFT, combined with state-of-the-art next-to-next-to-next-to-leading order (N\(^3\)LO) chiral
interactions, has led to remarkable agreement with experimental data across a range of nuclei~\cite{Elhatisari:2022zrb}.
Moreover, NLEFT has successfully described the structure of the
Hoyle state in \( ^{12}\text{C} \) \cite{Epelbaum:2011md,Epelbaum:2012qn}, \( \alpha \)-\( \alpha \) scattering processes
\cite{Elhatisari:2015iga}, geometric configurations of the \( ^{12}\text{C} \) spectrum \cite{Shen:2022bak}, nuclear
thermodynamics \cite{Lu:2019nbg}, clustering in hot dilute matter \cite{Ren:2023ued}, structure factors for hot
neutron matter \cite{Ma:2023ahg}, hyper-neutron matter \cite{Tong:2024jvs}, and hypernuclei \cite{Hildenbrand:2024ypw}.
In this work, we present a systematic \textit{ab initio} study of the \( p \)-shell beryllium isotopes using NLEFT
with the N\(^3\)LO interaction, including energies, radii, electromagnetic properties, and the geometric structures.

\section{Formalism}

We use the wavefunction matching method~\cite{Elhatisari:2022zrb} to mitigate the Monte Carlo sign problem 
associated with high-fidelity N${}^3$LO chiral interactions. This method unitarily transforms the original Hamiltonian, $H$, 
into a new high-fidelity Hamiltonian, $H'$, such that its wave  functions match those of a computationally simple Hamiltonian, $H^{\rm S}$, 
up to a given radius. This transformation ensures that the expansion in powers of the difference $H'-H^{\rm S}$ converges rapidly.
For more details see Ref.~\cite{Elhatisari:2022zrb}. For comparison, we also employ a simple SU(4)-symmetric interaction
for the study of $^{12}$C~\cite{Shen:2022bak}.

In the NLEFT framework~\cite{Lee:2008fa,Lahde:2019npb}, observables are calculated as
\begin{equation}\label{eq:O}
  \langle O \rangle = \lim_{\tau \to \infty}
  \frac{\langle \Psi_0| M^{L_{t}/2} \,  O  \, M^{L_{t}/2} | \Psi_0 \rangle}{\langle \Psi_0| M^{L_{t}} | \Psi_0 \rangle},
\end{equation}
where $\Psi_0$ is the initial wave function,
$M$ is the normal-ordered transfer matrix operator $:e^{- H^{\rm S} \, a_{t}}:$ with temporal lattice spacing $a_{t}$, 
and $L_{t}$ is the total number of temporal lattice steps.

To study the nuclear geometrical properties, we employ the pinhole algorithm~\cite{Elhatisari:2017eno} and 
its perturbative extension~\cite{Lu:2018bat}.
The method samples the positions of $A$-nucleons, denoted as $\mathbf{n}_i$, on the lattice (spin and isospin indices have been omited)
according to the following amplitude
\begin{equation}\label{eq:Z}
  Z = \langle \Psi_0| M^{L_{t}/2} \, \rho(\mathbf{n}_1,\mathbf{n}_2,\dots,\mathbf{n}_A) \, M^{L_{t}/2} | \Psi_0 \rangle,
\end{equation}
where $\rho(\mathbf{n}_1,\mathbf{n}_2,\dots,\mathbf{n}_A)$ is the
normal-ordered product of single-nucleon density operators
$\rho(\mathbf{n}_i) = a^\dagger(\mathbf{n}_i) a(\mathbf{n}_i)$.
Let $N_{\rm pin}$ represent the total number of sampled pinhole configurations. These configurations can be written as
\begin{equation}\label{eq:npin}
  \left\{ \mathbf{N}^{(k)} = \left( \mathbf{n}_1^{(k)}, \mathbf{n}_2^{(k)}, \dots, \mathbf{n}_A^{(k)} \right) \right\}_{k = 1}^{N_{\rm pin}}.
\end{equation}
where $N_{\rm pin}$ typically reaches several millions for the current study of beryllium isotopes. 
Each configuration then can be transformed into the $A$-nucleon center-of-mass (c.m.)
coordinate, $\mathbf{r}_i$, \cite{Elhatisari:2017eno}:
\begin{equation}\label{eq:rpin}
  \left\{ \mathbf{R}^{(k)} = \left( \mathbf{r}_1^{(k)}, \mathbf{r}_2^{(k)}, \dots, \mathbf{r}_A^{(k)} \right) \right\}_{k = 1}^{N_{\rm pin}}.
\end{equation}
To account for the finite nucleon size (0.84fm)~\cite{Lin:2021xrc}, a random Gaussian smearing is applied.

The quadrupole moment for a given configuration can then be calculated as
(the denominator required for normalization has been omitted)
\begin{equation}\label{eq:q}
  \langle Q \rangle = e \sum_{k=1}^{N_{\rm pin}} (-1)^{s_k} \sum_{i = 1}^Z \left[ r_i^{(k)} \right]^2 \left( 3 \cos^2\theta_i^{(k)} - 1 \right),
\end{equation}
with $s_k = 0$ or $1$ for the sign due to importance sampling to the absolute amplitude of $|Z|$~\cite{Lahde:2019npb,Elhatisari:2017eno}, and the summation is over protons.
The reduced transition probability is
\begin{equation}\label{eq:bex}
  \langle B(E\lambda;I_1\to I_2) \rangle = e^2 \sum_{\mu M_2} \bigg| \sum_{k=1}^{N_{\rm pin}} (-1)^{s_k} \sum_{i=1}^Z r_i^\lambda Y_{\lambda\mu}(\hat{\mathbf{r}}_i) \bigg|^2.
\end{equation}
The deformation parameters~\cite{ring2004nuclear} for a given pinhole configuration $(k)$ are
\begin{subequations}\label{eq:a2x}\begin{align}
  a_{20}^{(k)} &= \frac{4\pi}{3AR_0^2} \sqrt{\frac{ 5}{16\pi}} \sum_{i=1}^A \left( 3\left[ z_i^{(k)} \right]^2
  - \left[ r_i^{(k)} \right]^2 \right), \\
  a_{22}^{(k)} &= \frac{4\pi}{3AR_0^2} \sqrt{\frac{15}{32\pi}} \sum_{i=1}^A \left( \left[ x_i^{(k)} \right]^2
  - \left[ y_i^{(k)} \right]^2  + i x_i^{(k)}y_i^{(k)} \right),
\end{align}\end{subequations}
with $R_0 = 1.2$ fm $A^{1/3}$.
The Hill-Wheeler coordinates \cite{Hill:1952jb} $\beta, \gamma$ can be calculated with:
\begin{equation}\label{eq:btgm}
  a_{20}^{(k)} = \beta^{(k)} \cos \gamma^{(k)}, \quad
  a_{22}^{(k)} = \frac{1}{\sqrt{2}} \beta^{(k)} \sin \gamma^{(k)}.
\end{equation}
By a suitable rotation, the expectation value of $\langle xy \rangle$ vanishes.
The statistically average over $N_{\rm pin}$ configurations will give us a deformation distribution
for a given state of nucleus.
To distinguish it from single determinant deformation, we label it as $\beta_{\rm pin}$ and $\gamma_{\rm pin}$.

Note that the expressions in Eqs.~(\ref{eq:rpin}-\ref{eq:a2x}) have no explicit left $\langle L|$ and right
$|R\rangle$ states information, they are encoded in Eq.~(\ref{eq:Z}).
For transition obsevables~\eqref{eq:bex} a multichannel calculation with
 different bra and ket states will be performed.

\section{Results and discussion}

The N\(^3\)LO interaction described in Ref.~\cite{Elhatisari:2022zrb} and SU(4)-symmetric interaction
in Ref.~\cite{Shen:2022bak} are defined on lattices
with spatial spacings $a = 1.32$~fm and $1.64$~fm, respectively,
which correspond to momentum cutoffs $\Lambda = \pi / a \simeq 471$~MeV and $377$~MeV. 
Additionally, the temporal lattice spacings for these interactions are defined as $a_{t} = 0.20$~fm and $a_{t} = 0.55$~fm, respectively, 
for these interactions.
We perform our lattice calculation in a periodic cubic box with length
$L = 13.2$~fm for the N\(^3\)LO interaction and $14.8$~fm for the SU(4)-symmetric interaction.
See Ref.~\cite{SM} for the details on the configurations of the initial wave functions.

\begin{figure}[!htbp]
  \centering
  \includegraphics[trim={3cm 0 0 0},clip,width=0.5\textwidth]{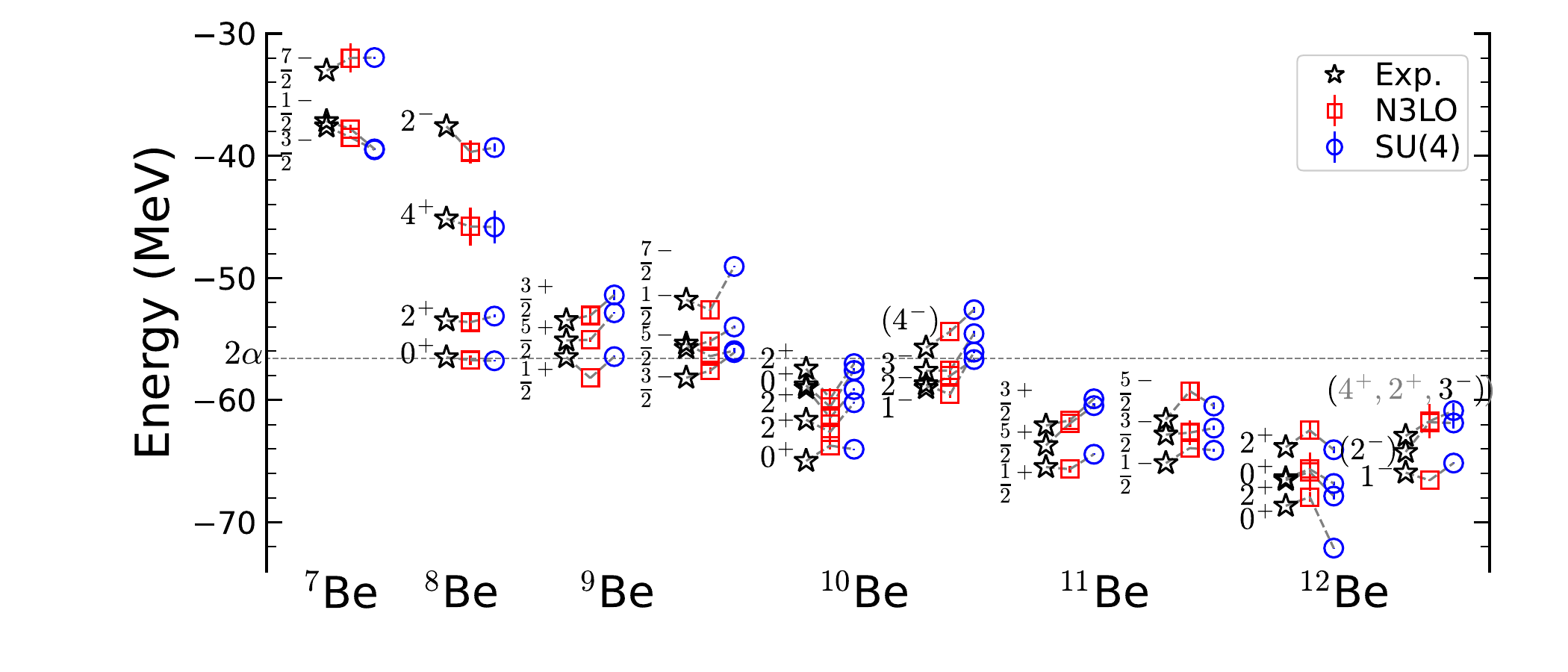}
  \caption{Low-lying spectrum from $^7$Be to $^{12}$Be calculated by NLEFT
  using N\(^3\)LO interaction~\cite{Elhatisari:2022zrb} and SU(4) interaction~\cite{Shen:2022bak}, compared to the
  data~\cite{Liu:2022zcn,Wang:2021xhn,Kelley:2017qgh,Kelley:2012qua,Tilley:2004zz,Tilley:2002vg}.
  The error bars correspond to one standard deviation errors include stochastic
  errors and uncertainties in the Euclidean time extrapolation.
  The two $\alpha$ threshhold is denoted by horizontal dashed line.}
  \label{fig1}
\end{figure}

Fig.~\ref{fig1} displays the low-lying energy spectra of \( ^{7}\text{Be} \) to \( ^{12}\text{Be} \) calculated using NLEFT
with the N\(^3\)LO interaction, compared to experimental data~\cite{Liu:2022zcn,Wang:2021xhn,Kelley:2017qgh,Kelley:2012qua,Tilley:2004zz,Tilley:2002vg}.
The trends of the theoretical predictions agree with the experimental results, affirming the effectiveness of the N\(^3\)LO interaction in
reproducing both ground and excited states of beryllium isotopes.
Numerical challenges such as Euclidean time extrapolation and finite volume effects become more pronounced for excited states~\cite{SM},
making it more difficult to maintain the same level of accuracy as for ground states. Addressing these challenges will require further optimization of 
computational algorithms and fine-tuning of three-body forces. Despite these challenges, the successful application of NLEFT to
beryllium isotopes underscores its potential for accurately capturing the complex dynamics of light nuclei.

It is noteworthy that the simple SU(4)-symmetric interaction~\cite{Shen:2022bak}
also provides an accurate description of most of the states, especailly \( ^{11}\text{Be} \). The ground state
of \( ^{11}\text{Be} \) has long posed a challenge to nuclear structure theory due to its inverted parity ordering,
where the \( 1/2^+ \) state lies below the \( 1/2^- \) state, contrary to shell-model predictions~\cite{Forssen:2004dk,Calci:2016dfb,Fortune:2018ees}.
Using only the SU(4)-symmetric interaction, we successfully reproduce this ordering with high precision: \( E(\frac{1}{2}^+) = -64.6(1) \) MeV
and \( E(\frac{1}{2}^-) = -64.1(1) \) MeV, compared to the experimental values of \( -65.5 \) MeV and \( -65.2 \) MeV, respectively.
This result underscores the importance of many-body correlations in achieving accurate nuclear structure descriptions.
\begin{table}[!tp]
    \centering
    \caption{Energies of states in $^7$Be and $^{10}$Be calculated by NLEFT
    that have not been identified by experiments.
    For $^7$Be some results from NCSM calculations~\cite{Vorabbi:2019imi} are listed for comparison.}
    \begin{tabular}{l|cc|c}
    \hline
                                    & NLEFT, N\(^3\)LO & NLEFT, SU(4) & NCSM \\
    \hline
    $~^{7}$Be , $(\frac{3}{2}^{+})$ & $-30.5(8)$  & $-29.9(3)$ & $-24.7$  \\
    $~^{7}$Be , $(\frac{1}{2}^{+})$ & $-28.8(1)$  & $-31.9(2)$ & $-27.8$  \\
    $~^{7}$Be , $(\frac{5}{2}^{+})$ & $-26.5(7)$  & $-26.5(1)$ &   \\
    \hline
    $~^{10}$Be, $A_1^{+}(3)$ & $-56.1(7)$ & $-58.4(9)$  & \\
    \hline
    \end{tabular}
    \label{tab:unknown}
\end{table}

By exploring various configurations, we identify states in beryllium isotopes that have not yet been observed experimentally, as listed in
Table~\ref{tab:unknown}. The existence of positive-parity states in \( ^7\text{Be} \) has been a subject of long-standing
debate~\cite{Piatti:2020gyp,He:2013ica}. Using shell-model wave functions
with one proton excited to the \( sd \) shell, our calculations yield lower energies compared to the NCSM
results~\cite{Vorabbi:2019imi}.

For \( ^{10}\text{Be} \), a three-channel \( 0^+ \) calculation with \textit{irrep} \( A_1^+ \) projection~\cite{Johnson:1982yq,Lu:2014xfa} reveals that
the ground state is a mixture of \( 1p_{3/2} \) and \( 1p_{1/2} \) channels, the second \( 0^+ \) state is predominant by \( sd \)-shell, and
the third state also comprises a mixture of \( 1p_{3/2} \) and \( 1p_{1/2} \). While the \( 0_2^+ \) state with \( sd \)-shell or
$\sigma$-orbital characteristics is well established~\cite{Kanada-Enyo:2018fjk,von2006nuclear}, the nature of the third \( A_1^{+} \) state
remains unclear. Although it could correspond to a \( 4^+ \) state, its calculated energy (\(-56 \sim -58\) MeV) is significantly
lower than the experimental \( 4_1^+ \) at \(-53.2\) MeV and the \(-49\) MeV obtained from similar \( J_z = 4 \) calculations.

\begin{figure}[!htbp]
  \includegraphics[trim={1cm 0.5cm 0 0.5cm},clip,width=0.4\textwidth]{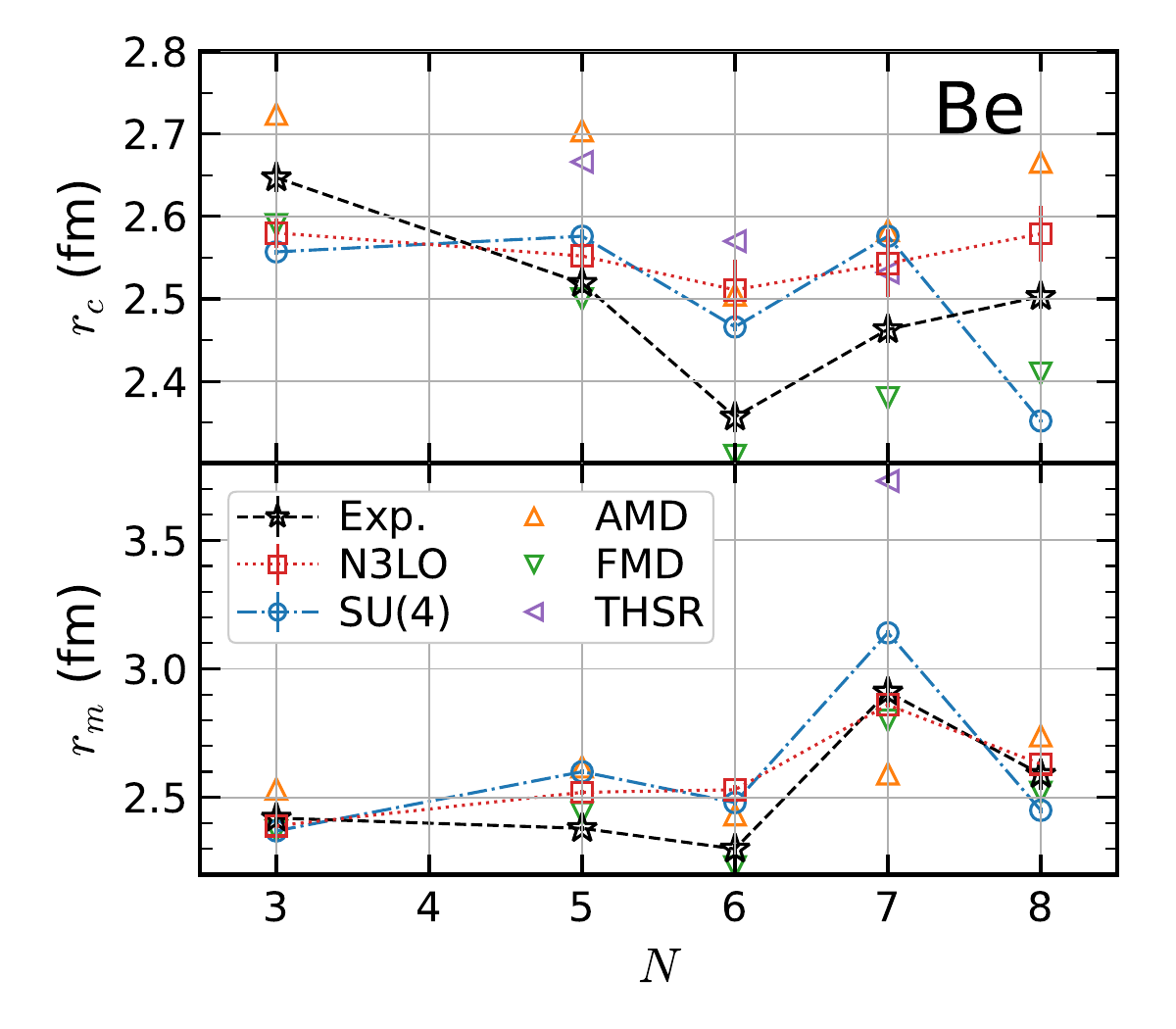}
  \caption{Radii from $^7$Be to $^{12}$Be calculated by NLEFT
  using N\(^3\)LO interaction~\cite{Elhatisari:2022zrb} and SU(4) interaction~\cite{Shen:2022bak}, compared to the
  data~\cite{Nortershauser:2008vp,Krieger:2012jx,Dobrovolsky:2019wzt,Tanihata:2013jwa}
  and other theoretical calculations \cite{Kanada-Enyo:2014sqb,Krieger:2012jx,Lyu:2014ewa,Lyu:2015ika,Lyu:2021ykp}.
  For Refs.~\cite{Kanada-Enyo:2014sqb,Lyu:2014ewa,Lyu:2021ykp} proton size, neutron size, and relativistic correction are added.
  (Upper panel) charge radii; (lower panel) point matter radii.
}
  \label{fig2}
\end{figure}

Fig.~\ref{fig2} displays the charge radii and point matter radii of beryllium isotopes with available experimental
data~\cite{Krieger:2012jx,Nortershauser:2008vp,Dobrovolsky:2019wzt,Tanihata:2013jwa}. Our theoretical results, compared to calculations
from AMD~\cite{Kanada-Enyo:2014sqb}, FMD~\cite{Krieger:2012jx}, and THSR~\cite{Lyu:2014ewa,Lyu:2015ika,Lyu:2021ykp}, agree with experimental
values within approximately 6\% and follow the same trend. Notably, the halo structure of \( ^{11}\text{Be} \) is accurately
reproduced using the N\(^3\)LO interaction.

\begin{table}[!tp]
    \centering
    \caption{The quadrupole moment and transition rates of Be isotopes calculated by NLEFT
    using the N\(^3\)LO interaction \cite{Elhatisari:2022zrb} and SU(4) interaction \cite{Shen:2022bak},
    in comparison with experiment.
    Units for $Q$ and $m(E0)$ are $e$fm$^2$, for $B(E1)$ are $e^2$fm$^2$, and for $B(E2)$ are $e^2$fm$^4$.}
    \begin{tabular}{ll|cc|c}
    \hline
                    &               & SU(4) & N\(^3\)LO & Exp. \\
    \hline
$^{7}$Be  & $E2,\frac{3}{2}^{-}\to \frac{1}{2}^{-}$ & 16.0(2)   & 15.2(5)    & 26(6)(3) \cite{Henderson:2019ubp} \\
\hline
$^{9}$Be  & $Q(\frac{3}{2}^{-})$                    & 7.3(1)    & 7.4(1.0)   & 5.29(4) \cite{Stone:2016bmk}  \\
          & $E1,\frac{1}{2}^{+}\to \frac{3}{2}^{-}$ & 0.131(3)  & 0.060(15)  & 0.136(2) \cite{Arnold:2011nv} \\
          & $E1,\frac{5}{2}^{+}\to \frac{3}{2}^{-}$ & 0.045(14) & 0.049(5)   & 0.010(8) \cite{Tilley:2004zz} \\
          & $E2,\frac{5}{2}^{-}\to \frac{3}{2}^{-}$ & 35.7(1.8) & 27.8(1.9)  & 27.1(2.0) \cite{Tilley:2004zz} \\
          & $E2,\frac{7}{2}^{-}\to \frac{3}{2}^{-}$ & 11.6(2.5) & 5.3(8)     & 9.5(4.1) \cite{Tilley:2004zz} \\
\hline
$^{10}$Be & $E1,3_1^{-}\to 2_1^{+}$                 & 0.026(2)  &  0.004(3)  & 0.009(1) \cite{Tilley:2004zz} \\
          & $E2,2_1^{+}\to 0_1^{+}$                 & 10.6(4)   & 8.5(9)     & 9.2(3) \cite{McCutchan:2009th}\\
\hline
$^{11}$Be & $E1,\frac{1}{2}^{-}\to \frac{1}{2}^{+}$ & 0.023(3)  & 0.038(3)   & 0.102(2) \cite{Kwan:2014dha} \\
\hline
$^{12}$Be & $E1,0_{1}^{+}\to 1_{1}^{-}$             & 0.049(2)  & 0.056(26)  & 0.051(13) \cite{Iwasaki:2000gp}     \\
          & $E2,2_{1}^{+}\to 0_{1}^{+}$             & 7.8(1.1)  & 9.0(3.1)   & 14.2(1.0)(2.0) \cite{Morse:2018ojw} \\
    \hline
    \end{tabular}
    \label{tab:trans}
\end{table}

We present the calculated quadrupole moments and transition rates for \( ^7\text{Be} \) to \( ^{12}\text{Be} \) in Table~\ref{tab:trans}. These
transition calculations are challenging due to slow convergence in Euclidean time and complex multichannel dynamics~\cite{Caprio:2022mkg}.
To address this, Euclidean time extrapolation has been employed, see~\cite{SM}. Our results generally agree with the
experimental data, with deviations observed in some cases. Given that electromagnetic observables are highly sensitive to nuclear
geometric structures, achieving precise reproduction is inherently ambitious.
Many theoretical studies (see Table~\ref{tab:trans} and \cite{Heng:2016umo,Pervin:2007sc,Pastore:2012rp,Kanada-Enyo:2015knx,Pieper:2002ne,Arai:2003jm,Myo:2023alz,Descouvemont:2020kwq,Orce:2012yc,Dan:2021htj,Dufour:2010dmf,Kanada-Enyo:2003fhn}) have examined these electromagnetic properties, and we compare their findings in~\cite{SM}.
Additionally, we are
developing a new method that combines second-order perturbative Monte Carlo~\cite{Lu:2021tab} with a trimmed sampling algorithm~\cite{Hicks:2022ovs}
to study transitions involving second \( 0^+ \) and \( 2^+ \) states using the N\(^3\)LO interaction.

\begin{figure*}[!htbp]
  \includegraphics[width=0.8\textwidth]{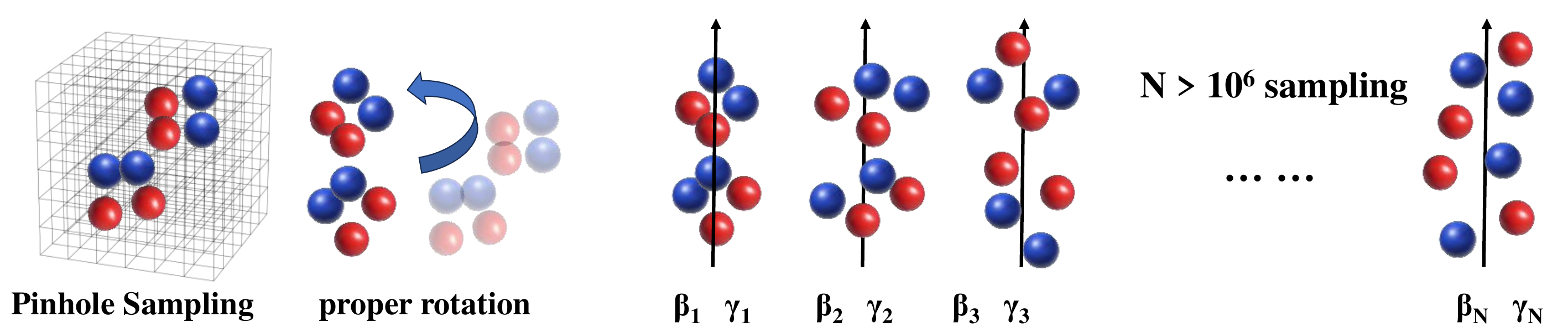}
  \includegraphics[trim={1cm 0 0 0},clip,width=1.0\textwidth]{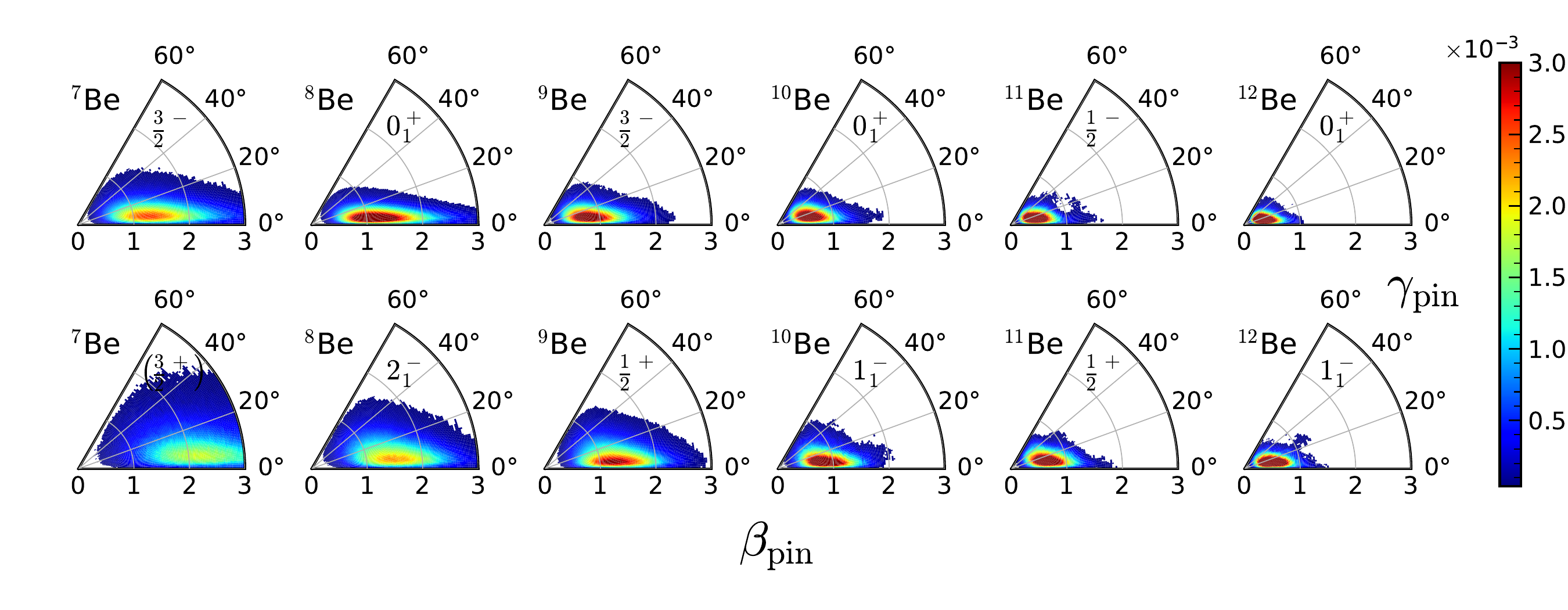}
  \caption{Probability distribution of the deformation parameter $\beta_{\rm pin}$
  and $\gamma_{\rm pin}$ in Eq.~(\ref{eq:btgm}) through sampling of pinhole configuration by NLEFT using N\(^3\)LO interaction,
  with red color represent a higher probability and blue color lower. The third component of total spin is fixed at $J_z = J$.}
  \label{fig3}
\end{figure*}

Recent experimental advancements, such as the collective-flow-assisted nuclear shape-imaging technique introduced by STAR~\cite{STAR:2024wgy}, have provided unprecedented insights into the shapes of atomic nuclei, highlighting the need for complementary theoretical approaches. In Fig.~\ref{fig3}, we present the probability distributions of the deformation parameters $\beta_{\rm pin}$ and $\gamma_{\rm pin}$ for the beryllium isotopes. This model-independent analysis offers a statistical representation of the relative positions of all nucleons, distinguishing it from traditional energy surface plots based on single Slater determinants.
Our results demonstrate that the occupation of different valence neutron orbitals—specifically $\pi$- or $\sigma$-orbitals significantly alters the nuclear shape. In particular, valence neutrons occupying $\sigma$-orbitals lead to more prolate deformations, whereas the occupation of $\pi$-orbitals results in more spherical shapes. These findings are consistent with the nuclear molecular framework \cite{Kanada-Enyo:2012yif, Yoshida:2013dwa, von2006nuclear, Carlson:2014vla}.

\begin{figure}[!htbp]
  \includegraphics[width=0.5\textwidth]{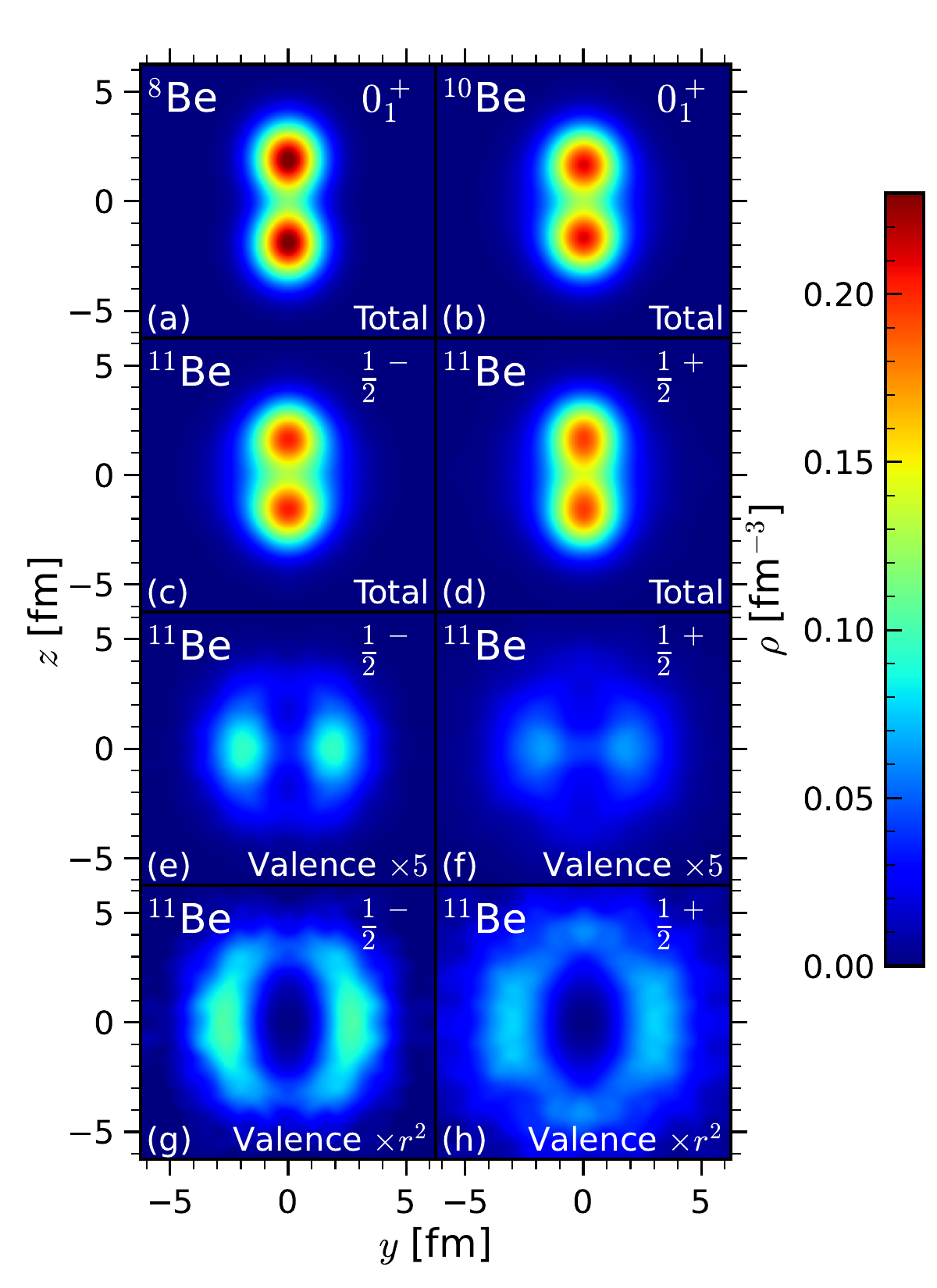}
  \caption{Intrinsic density at $x = 0$ plane of selected states of beryllium isotopes obtained by NLEFT using N\(^3\)LO interaction. The third component of total spin is fixed at $J_z = J$.}
  \label{fig4}
\end{figure}

Finally, we present intrinsic density plots of selected states in beryllium isotopes in Fig.~\ref{fig4}. To obtain these intrinsic densities,
we adopt the strategy from Ref.~\cite{Shen:2022bak}, which groups the closest two protons and two neutrons and randomly aligns the
clusters along the $\pm z$-axis. This method ensures a balanced representation of nuclear shapes, avoiding the overemphasis of any single
axis that occurs when aligning configurations based on the principal axis \cite{Wiringa:2000gb, SM}.

Panels~(a)–(d) display the total density for \( ^8\text{Be} \), \( ^{10}\text{Be} \), and \( ^{11}\text{Be} \) (\(1/2^-\) and \(1/2^+\) states).
\( ^8\text{Be} \) clearly shows a strong two-alpha cluster structure as expected. Adding valence neutrons in \( ^{10}\text{Be} \) and
\( ^{11}\text{Be} \) slightly diminishes the cluster formation while enhancing the neck region between clusters. Comparing the \(1/2^-\) and
\(1/2^+\) states of \( ^{11}\text{Be} \), we observe significantly different shapes: the \( \pi \)-orbital occupation results in a more
rounded nucleus, whereas the \( \sigma \)-orbital induces a pronounced prolate deformation, consistent with nuclear molecular
dynamics in other studies \cite{Kanada-Enyo:2012yif,Yoshida:2013dwa,von2006nuclear,Carlson:2014vla}.

Panels~(e)–(h) illustrate the valence neutron densities, scaled by a constant factor of 5 or by \( r^2 \). In panel~(e), the \( \pi \)-orbital
in \( ^{11}\text{Be} \) naturally emerges from the N\(^3\)LO interaction, displaying a distinct distribution. For the \(1/2^+\)
state in panel~(f), one neutron occupies the \( \sigma \)-orbital, reducing the \( \pi \)-orbital density. Applying an \( r^2 \)
scaling in panels~(g) and (h) reveals the large spatial extension of the last neutron in the \(1/2^+\) state, characteristic of a halo nucleus.
This extended distribution aligns with other models~\cite{Kanada-Enyo:2002owv,von2006nuclear}, showing enhanced density around \( r \sim 0 \) and
along the $\pm z$-axis, alongside shell-model \( sd \) characteristics indicative of a halo structure.

The concept of nuclear molecular orbitals has been extensively discussed in cluster models \cite{von2006nuclear}, but it remains less straightforward in the context of \textit{ab initio} calculations. The primary challenge lies in identifying the clusters and valence particles within the full many-body correlated wave function $\Psi(\mathbf{r}_1,\mathbf{r}_2, \dots ,\mathbf{r}_A)$. The current work offers new insights into this task. By performing Monte Carlo sampling of the many-body density operator, the pinhole algorithms provide configurations of the $A$-particle coordinates in space.
Furthermore, by grouping the closest $2$ protons and $2$ neutrons together, the remaining particles are automatically categorized as valence particles. 
In the Supplemental Material, we describe in detail the grouping 
algorithm used, and we also explicitly construct a simple model 
for the $\pi$ and $\sigma$ molecular orbitals that is able to 
reproduce the {\it ab initio} nucleonic densities seen for the 
\(3/2^-\) and \(1/2^+\) states of $^9$Be as well as  the ground states 
of $^{10}$Be, $^{11}$Be, and $^{12}$Be \cite{SM}.
This effectively reveals the nuclear molecular orbitals directly from the full wave function $\Psi(\mathbf{r}_1,\mathbf{r}_2, \dots ,\mathbf{r}_A)$,
offering an \textit{ab initio} description of these orbitals.

\section{Summary and discussion}

We have systematically studied the \( p \)-shell beryllium isotopes using nuclear lattice effective field theory (NLEFT) with both the
N\(^3\)LO interaction~\cite{Elhatisari:2022zrb} and a simple SU(4)-symmetric interaction~\cite{Shen:2022bak}. Our calculations for the low-lying
spectra, radii, and electromagnetic observables show good agreement with experimental data.
We have investigated the halo structure of \( ^{11}\text{Be} \), the geometric differences between negative- and positive-parity states,
and intrinsic density distributions.
By identifing clusters and valence neutrons from the pinhole algorithm,
nuclear molecular orbitals, e.g. $\pi$- and $\sigma$-orbitals, emerge naturally.
These findings demonstrate the efficiency of NLEFT in capturing the intricate dynamics of
light nuclei, highlighting the potential of unified \textit{ab initio} approaches in elucidating complex
nuclear behaviors such as the nature and details of the molecular orbitals in nuclei with clustering.


\section{Acknowledgments}
We are grateful for discussions with Jie Meng and the members of the NLEFT Collaboration.
This work is supported in part by the European
Research Council (ERC) under the European Union's Horizon 2020 research
and innovation programme (ERC AdG EXOTIC, grant agreement No. 101018170),
by DFG and NSFC through funds provided to the
Sino-German CRC 110 ``Symmetries and the Emergence of Structure in QCD" (NSFC
Grant No.~11621131001, DFG Grant No.~TRR110).
The work of SS was supported in by the National Natural Science Foundation of China under
Grants No.12435007.
The work of UGM was supported in by the CAS President's International
Fellowship Initiative (PIFI) (Grant No.~2025PD0022).
The work of SE is supported in part by the Scientific and Technological Research Council of Turkey (TUBITAK project no. 123F464).
The work of DL is supported in part by the U.S. Department of Energy (Grants No. DE-SC0013365, 
No. DE-SC0023658, No. DE-SC0024586), U.S. National Science Foundation (Grant No. PHY-2310620),
and the Nuclear Computational Low-Energy Initiative (NUCLEI) SciDAC project.
The authors gratefully acknowledge the Gauss Centre for Supercomputing e.V. (www.gauss-centre.eu)
for funding this project by providing computing time on the GCS Supercomputer JUWELS
at J\"ulich Supercomputing Centre (JSC).

\putbib[bref-be]

\end{bibunit}

\clearpage

\beginsupplement
\begin{bibunit}[apsrev4-2]
%
%
%
%
%
%
%


\begin{appendix}
\begin{onecolumngrid}

\renewcommand{\thefigure}{S\arabic{figure}}
\renewcommand{\thetable}{S\arabic{table}}
\renewcommand{\theequation}{S\arabic{equation}}
\setcounter{figure}{0}
\setcounter{equation}{0}
\setcounter{table}{0}

\section*{Supplemental Material}

We give further information on the initial wave configuarations,
table of numerical values, various numerical checks
such as Euclidean time extrapolation, box size, the breaking of cubic irrepresentation, and more intrinsic density information.
All the results below are obtained by full N\(^3\)LO interaction if not stated differently.

\subsection{Euclidean time extrapolation}

We use one exponential formulas for the extrapolation of observables.
For the nonperturbative energy $E_0$ of the channel $i$, this reads:
\begin{equation}\label{eq:extra-e}
  E_0^{(i)} (\tau) = \frac{E_\infty^{(i)} + (E_\infty^{(i)} + d^{(i)}) c^{(i)} e^{-d^{(i)}\tau}}{1+c^{(i)}e^{-d^{(i)}\tau}},
\end{equation}
with $E_\infty, c,$ and $d$ the fit parameters.

For other observables, such as the  energy and radius, we use
\begin{equation}\label{eq:extra-o1}
  O^{(i)} (\tau) = \frac{O_\infty^{(i)} + O_1^{(i)} e^{-d^{(i)}\tau/2} + O_2^{(i)} e^{-d^{(i)}\tau}}{1+c^{(i)}e^{-d^{(i)}\tau}},
\end{equation}
with $O_\infty, O_1,$ and $O_2$ as new fit parameters, while the parameters $c$ and $d$
will be fitted together with Eq.~(\ref{eq:extra-e}).
For observables with mixed channels such as transitions, the extrapolation formula is extended to
\begin{equation}\label{eq:extra-o2}
  O^{(ij)} (\tau) = \frac{O_\infty^{(ij)} + O_1^{(i)} e^{-d^{(i)}\tau/2} + O_1^{(j)} e^{-d^{(j)}\tau/2}}{
  \left[ 1+c^{(i)}e^{-d^{(i)}\tau} \right]^{1/2} \left[ 1+c^{(j)}e^{-d^{(j)}\tau} \right]^{1/2}}.
\end{equation}
Similarly to the above, theparameters $c$ and $d$ will be fitted together with Eq.~(\ref{eq:extra-e}).

\begin{figure}[!htbp]
  \centering
  \includegraphics[width=0.49\textwidth]{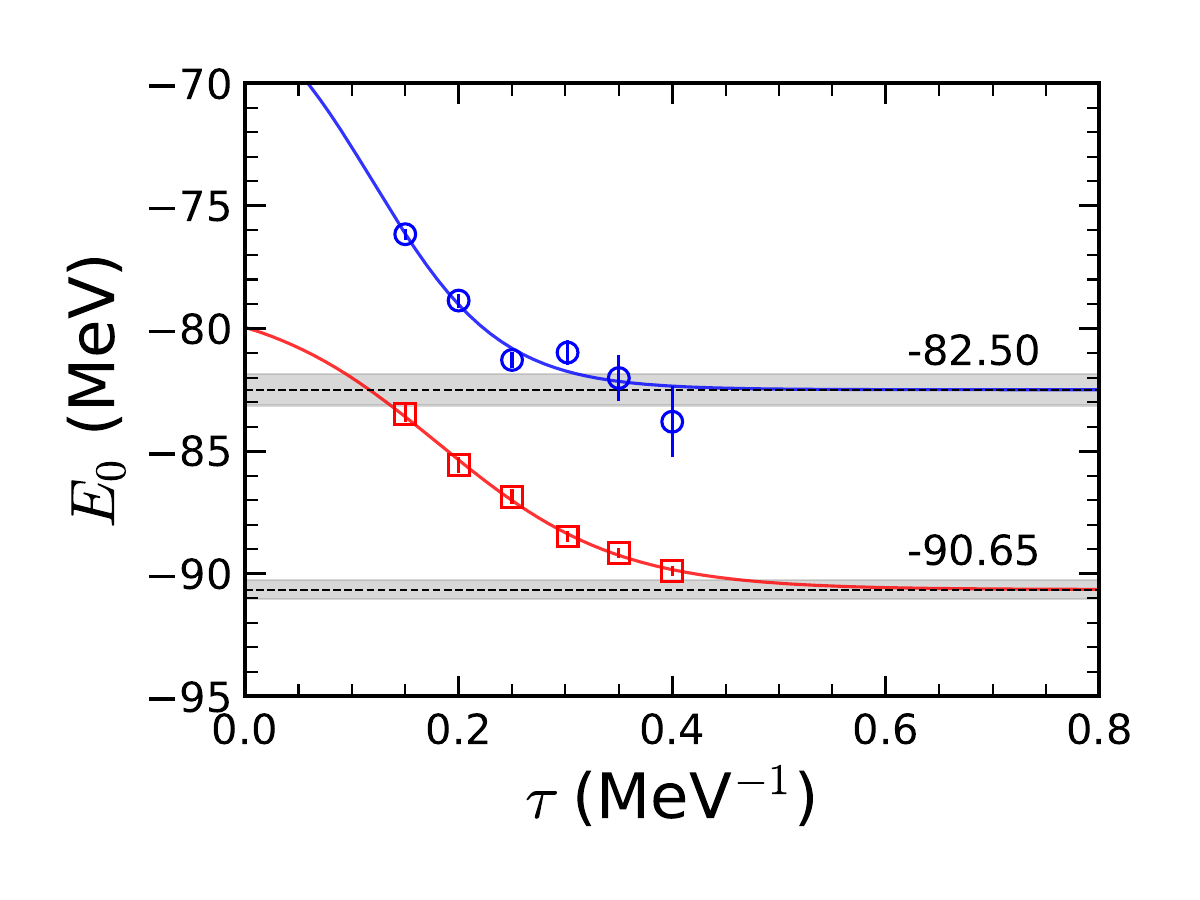}
  \includegraphics[width=0.49\textwidth]{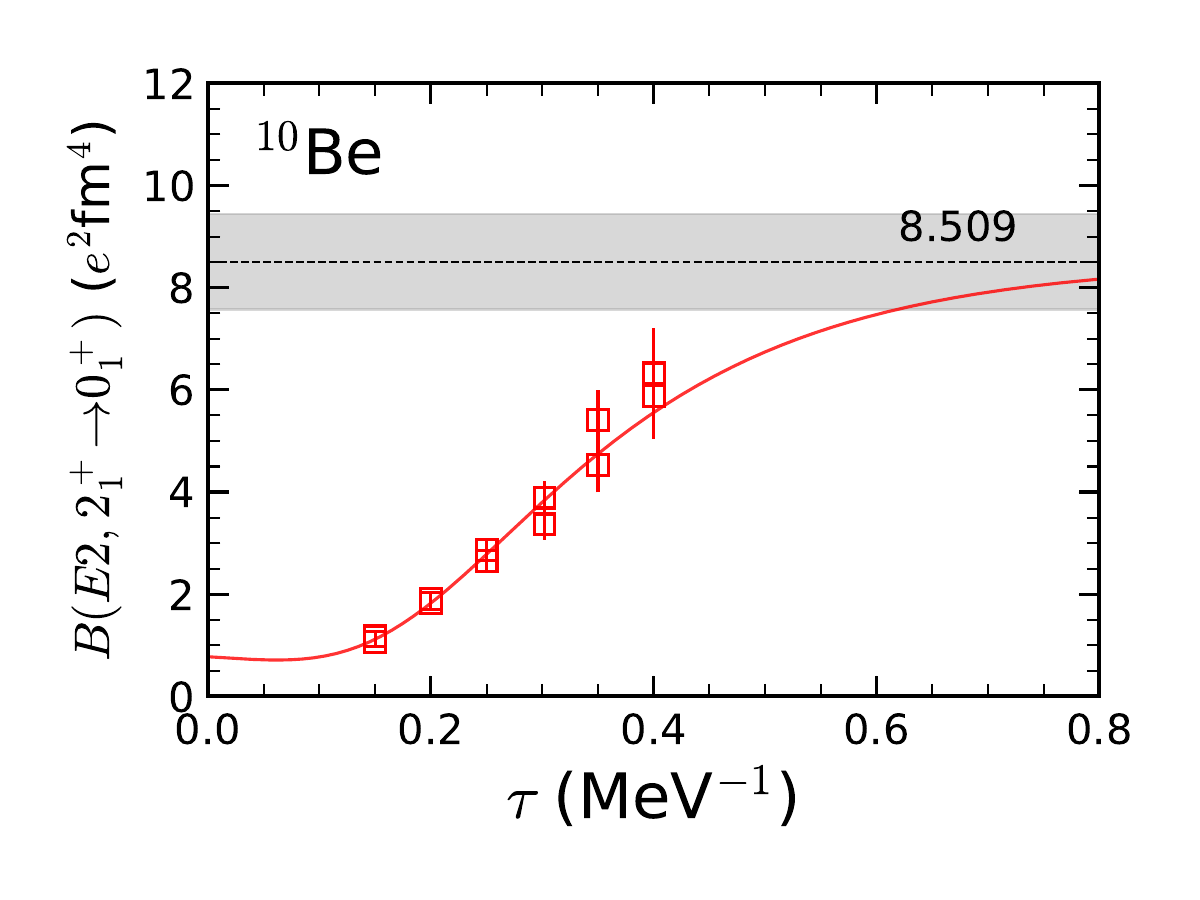}
  \caption{(Left panel) Extrapolation of the nonperturbative energy of the
  $0_1^+$ (red squares and lines) and $2_1^+$ (blue circles and lines) states
  in $^{10}$Be calculated in  NLEFT. (Right panel) Extrapolation of $B(E2,2_1^+\to 0_1^+)$.
  The gray bands indicate the error of the parameters $E_\infty$ and $O_\infty$.}
  \label{fig:extra}
\end{figure}

As an example, in Fig.~\ref{fig:extra}, we show the Euclidean time extrapolation of  the nonperturbative
energies and transition $B(E2,2_1^+\to 0_1^+)$ in $^{10}$Be.
The sign problem and computational complexity increase dramatically as the projection time $\tau$ increases,
and therefore we will limit our resources to a reasonable range where the simulation gives reliable results.
A multi-exponential extrapolation formula is not chosen to avoid overfitting.

\subsection{Finite volume effect}

We use two states that have a large spatial extension to show the finite volume effect: the first is
the high-lying resonance $\frac{3}{2}^+$ in $^7$Be and the second is the neutron halo state
of $\frac{1}{2}^+$ in $^{11}$Be.

\begin{table}[h]
\begin{tabular}{|c|c|c|}
\hline
$L$ [fm] & $^7$Be, $E(\frac{3}{2}^+)$ [MeV] & $^{11}$Be, $E(\frac{1}{2}^+)$ [MeV] \\
\hline
13.2  &  $-31.6(6)$    & $65.4(3)$ \\
15.7  &  $-28.2(1.8)$  & $65.3(6)$ \\
\hline
\end{tabular}
\caption{Selected energies calculated at different box sizes $L$. The error bars indicate stochastic errors.
\label{tab:fv}}
\end{table}

In Table~\ref{tab:fv} we show the results calculated at $L_t = 500$ for different box sizes.
It can be seen for bound halo structure, the  box size $L = 13.2$ fm issufficient; while for
unbound resonances, a mild finite volmue effect can be seen. This can be improved in the
future by better methods to address such type of states.

\subsection{Rotational irrepresentation on lattice}

Rotational symmetry including spin-$1/2$ particles on the lattice can be expressed
by 8 irreducible representations (\textit{irreps}): $A_1, A_2, E, T_1, T_2, G_1, G_2, H$ \cite{Johnson:1982yq,Lu:2014xfa}.
Here we check how much the results are affected by the breaking of different \textit{irreps}.
We take the example of $2_1^+$ in $^{10}$Be, with initial wave
function of $J_z^\pi = 2^+$ shell-model wave function and can be both \textit{irreps} of $E^+$ and $T_2^+$.
For energies the results are obtained at $L_t = 200$,
and for $B(E2)$ at $L_t = 150$ due to large computational complexity.
The comparison of two \textit{irreps} $E^+$ and $T_2^+$ are listed in Table~\ref{tab:irrep}

\begin{table}[h]
\begin{tabular}{|c|c|c|}
\hline
\textit{irrep} & $E(2_1^+)$ [MeV] &  $B(E2,2_1^+\to 0_1^+)$ [$e^2$fm$^4$] \\
\hline
$E^+$  &  $-61.6(3)$  & $2.72(49)$ \\
$T_2^+$&  $-58.4(7)$  & $1.95(61)$ \\
\hline
\end{tabular}
\caption{Energy and $B(E2)$ transition of $2_1^+$ in $^{10}$Be for
different \textit{irreps}. The error bars indicate stochastic errors.
\label{tab:irrep}}
\end{table}

It can be seen the error due to breaking of different \textit{irreps} is about 1.6~MeV in the energy
and 0.38~$e^2$fm$^4$ in the $B(E2)$ transitions.
The results presented in the paper are obtained using $J_z = 2$ initial wave function
without doing any \textit{irrep} projection, therefore  the two \textit{irreps} are automatically averaged.

\subsection{Initial wave function}

A single Slater determinant composed of shell-model wave functions
has been used as the initial wave function $|\Psi_0\rangle$ throughout this paper.
Comparing with cluster wave function, the main advantage of using shell-model wave function is
that a good quantum number $J_z$ can be constructed and this helps to identify the
total spin of the final state.
With cluster wave function a projection into \textit{irreps} is necessary and this often causes a severe sign problem.

In Table~\ref{tab:iniwav} we list the configurations we use for the initial wave function for different states.
Note  that the first 2 protons and neutrons always occupy the $1s_{1/2}$ orbital and therefore they are not listed.
In some cases the order is not natural but is chosen for a numerical advantage (e.g. smaller error or lower energy),
and since this is only a initial wave function, such change will not modify the final results.
Furthermore, as there is no requirement of exact orthogonality, the single-particle wave function
for each nucleon can be slightly adjusted to achieve a faster convergence such as using different
harmonic oscillator strength.

\pagebreak

\begin{longtable}{|l|cc|cccccc|c|}
    \caption{Proton ($p_i, i = 3, 4$) and neutron ($n_i, i = 3, \dots, N$) configuration of initial wave functions in NLEFT calculation for beryllium isotopes.
    The notation $l_{j}^{j_z}$ is used to denote the three quantum numbers $l,j,j_z$ (orbital angular momentum, total angular momentum, projection of total angular momentum onto $z$-axis).
    The radial quantum number $n$ is omitted to save space, for all $p$ and $d$ orbitals $n = 1$ ,
    while for all $s$ orbitals $n = 2$.
    Projection of the total spin onto  the$z$-axis and parity ($J_z^\pi$) are listed in the last column.}\\
    \hline
    & $p_3$ & $p_4$ & $n_3$ & $n_4$ & $n_5$ & $n_6$ & $n_7$ & $n_8$ & $J_z^\pi$ \\
    \hline
$~^{7}$Be ,$\frac{3}{2}^{-}$  &$p_{3/2}^{+1/2}$&$p_{3/2}^{-1/2}$&$p_{3/2}^{+3/2}$&                 &  &  & & & $\frac{3}{2}^{-}$ \\
$~^{7}$Be ,$\frac{1}{2}^{-}$  &$p_{3/2}^{+1/2}$&$p_{3/2}^{-1/2}$&$p_{1/2}^{+1/2}$&                 &  &  & & & $\frac{1}{2}^{-}$ \\
$~^{7}$Be ,$\frac{7}{2}^{-}$  &$p_{1/2}^{+1/2}$&$p_{3/2}^{+3/2}$&$p_{3/2}^{+3/2}$&                 &  &  & & & $\frac{7}{2}^{-}$ \\
$~^{7}$Be ,$(\frac{1}{2})^{+}$&$s_{1/2}^{+1/2}$&$p_{3/2}^{-1/2}$&$p_{3/2}^{+1/2}$&                 &  &  & & & $\frac{1}{2}^{+}$ \\
$~^{7}$Be ,$(\frac{3}{2})^{+}$&$p_{3/2}^{+1/2}$&$s_{1/2}^{+1/2}$&$p_{3/2}^{+1/2}$&                 &  &  & & & $\frac{3}{2}^{+}$ \\
$~^{7}$Be ,$(\frac{5}{2})^{+}$&$s_{1/2}^{+1/2}$&$p_{3/2}^{+3/2}$&$p_{3/2}^{+1/2}$&                 &  &  & & & $\frac{5}{2}^{+}$ \\
    \hline
$~^{8}$Be ,$0_{1}^{+}$        &$p_{3/2}^{+1/2}$&$p_{3/2}^{-1/2}$&$p_{3/2}^{+1/2}$&$p_{3/2}^{-1/2}$&  &  & & & $          0^{+}$ \\
$~^{8}$Be ,$2_{1}^{+}$        &$p_{3/2}^{+1/2}$&$p_{3/2}^{+3/2}$&$p_{3/2}^{+1/2}$&$p_{3/2}^{-1/2}$&  &  & & & $          2^{+}$ \\
$~^{8}$Be ,$4_{1}^{+}$        &$p_{3/2}^{+1/2}$&$p_{3/2}^{+3/2}$&$p_{3/2}^{+1/2}$&$p_{3/2}^{+3/2}$&  &  & & & $          4^{+}$ \\
$~^{8}$Be ,$2_{1}^{-}$        &$p_{3/2}^{+1/2}$&$s_{1/2}^{-1/2}$&$p_{3/2}^{+1/2}$&$p_{3/2}^{-1/2}$&  &  & & & $          0^{-}$ \\
    \hline
$~^{9}$Be ,$\frac{3}{2}^{-}$  &$p_{3/2}^{+1/2}$&$p_{3/2}^{-1/2}$&$p_{3/2}^{+1/2}$&$p_{3/2}^{-1/2}$&$p_{3/2}^{+3/2}$&  & & & $\frac{3}{2}^{-}$ \\
$~^{9}$Be ,$\frac{1}{2}^{+}$  &$p_{3/2}^{+1/2}$&$p_{3/2}^{-1/2}$&$p_{3/2}^{+1/2}$&$p_{3/2}^{-1/2}$&$s_{1/2}^{+1/2}$&  & & & $\frac{1}{2}^{+}$ \\
$~^{9}$Be ,$\frac{5}{2}^{-}$  &$p_{3/2}^{+3/2}$&$p_{3/2}^{-1/2}$&$p_{3/2}^{+1/2}$&$p_{3/2}^{-1/2}$&$p_{3/2}^{+3/2}$&  & & & $\frac{5}{2}^{-}$ \\
$~^{9}$Be ,$\frac{1}{2}^{-}$  &$p_{3/2}^{+1/2}$&$p_{3/2}^{-1/2}$&$p_{3/2}^{+1/2}$&$p_{3/2}^{-1/2}$&$p_{1/2}^{+1/2}$&  & & & $\frac{1}{2}^{-}$ \\
$~^{9}$Be ,$\frac{5}{2}^{+}$  &$p_{3/2}^{+1/2}$&$p_{3/2}^{-1/2}$&$p_{3/2}^{+1/2}$&$p_{3/2}^{-1/2}$&$d_{5/2}^{+5/2}$&  & & & $\frac{5}{2}^{+}$ \\
$~^{9}$Be ,$\frac{3}{2}^{+}$  &$p_{3/2}^{+3/2}$&$p_{3/2}^{-1/2}$&$p_{3/2}^{+1/2}$&$p_{3/2}^{-1/2}$&$d_{5/2}^{+1/2}$&  & & & $\frac{3}{2}^{+}$ \\
$~^{9}$Be ,$\frac{7}{2}^{-}$  &$p_{3/2}^{+1/2}$&$p_{3/2}^{+3/2}$&$p_{3/2}^{+1/2}$&$p_{3/2}^{-1/2}$&$p_{3/2}^{+3/2}$&  & & & $\frac{7}{2}^{-}$ \\
    \hline
$~^{10}$Be,$0_{1}^{+}$        &$p_{3/2}^{+1/2}$&$p_{3/2}^{-1/2}$&$p_{3/2}^{+1/2}$&$p_{3/2}^{-1/2}$&$p_{3/2}^{+3/2}$&$p_{3/2}^{-3/2}$& & & $0^{+}$ \\
$~^{10}$Be,$2_{1}^{+}$        &$p_{3/2}^{+1/2}$&$p_{3/2}^{+3/2}$&$p_{3/2}^{+1/2}$&$p_{3/2}^{-1/2}$&$p_{3/2}^{+3/2}$&$p_{3/2}^{-3/2}$& & & $2^{+}$ \\
$~^{10}$Be,$2_{2}^{+}$        &$p_{3/2}^{+1/2}$&$p_{3/2}^{-1/2}$&$p_{3/2}^{+1/2}$&$p_{3/2}^{-1/2}$&$p_{3/2}^{+3/2}$&$p_{1/2}^{+1/2}$& & & $2^{+}$ \\
$~^{10}$Be,$1_{1}^{-}$        &$p_{3/2}^{+1/2}$&$p_{3/2}^{-1/2}$&$p_{3/2}^{+1/2}$&$p_{3/2}^{-1/2}$&$p_{1/2}^{+1/2}$&$s_{1/2}^{+1/2}$& & & $1^{-}$ \\
$~^{10}$Be,$0_{2}^{+}$        &$p_{3/2}^{+1/2}$&$p_{3/2}^{-1/2}$&$p_{3/2}^{+1/2}$&$p_{3/2}^{-1/2}$&$s_{1/2}^{+1/2}$&$s_{1/2}^{-1/2}$& & & $0^{+}$ \\
$~^{10}$Be,$2_{1}^{-}$        &$p_{3/2}^{+1/2}$&$p_{3/2}^{-1/2}$&$p_{3/2}^{+1/2}$&$p_{3/2}^{-1/2}$&$p_{3/2}^{+3/2}$&$s_{1/2}^{+1/2}$& & & $2^{-}$ \\
$~^{10}$Be,$3_{1}^{-}$        &$p_{3/2}^{+1/2}$&$p_{3/2}^{-1/2}$&$p_{3/2}^{+1/2}$&$d_{5/2}^{+5/2}$&$p_{3/2}^{+3/2}$&$p_{3/2}^{+3/2}$& & & $3^{-}$ \\
$~^{10}$Be,$A_1^+(3)$         &$p_{3/2}^{+1/2}$&$p_{3/2}^{-1/2}$&$p_{3/2}^{+1/2}$&$p_{3/2}^{-1/2}$&$p_{1/2}^{+1/2}$&$p_{1/2}^{-1/2}$& & & $0^{+}$ \\
$~^{10}$Be,$2_{3}^{+}$        &$p_{3/2}^{+1/2}$&$p_{3/2}^{-1/2}$&$p_{3/2}^{+1/2}$&$s_{1/2}^{-1/2}$&$p_{3/2}^{+3/2}$&$s_{1/2}^{+1/2}$& & & $2^{+}$ \\
$~^{10}$Be,$(4_{1}^{-})$      &$p_{3/2}^{+1/2}$&$p_{3/2}^{-1/2}$&$p_{3/2}^{+1/2}$&$p_{3/2}^{-1/2}$&$p_{3/2}^{+3/2}$&$d_{5/2}^{+5/2}$& & & $4^{-}$ \\
    \hline
$~^{11}$Be $\frac{1}{2}^{+}$  &$p_{3/2}^{+1/2}$&$p_{3/2}^{-1/2}$&$p_{3/2}^{+1/2}$&$p_{3/2}^{-1/2}$&$p_{3/2}^{+3/2}$&$p_{3/2}^{-3/2}$&$s_{1/2}^{+1/2}$& & $\frac{1}{2}^{+}$ \\
$~^{11}$Be $\frac{1}{2}^{-}$  &$p_{3/2}^{+1/2}$&$p_{3/2}^{-1/2}$&$p_{3/2}^{+1/2}$&$p_{3/2}^{-1/2}$&$p_{3/2}^{+3/2}$&$p_{3/2}^{-3/2}$&$p_{1/2}^{+1/2}$& & $\frac{1}{2}^{-}$ \\
$~^{11}$Be $\frac{5}{2}^{+}$  &$p_{3/2}^{+1/2}$&$p_{3/2}^{-1/2}$&$p_{3/2}^{+1/2}$&$p_{3/2}^{-1/2}$&$p_{3/2}^{+3/2}$&$p_{3/2}^{-3/2}$&$d_{5/2}^{+5/2}$& & $\frac{5}{2}^{+}$ \\
$~^{11}$Be $\frac{3}{2}^{-}$  &$p_{3/2}^{+3/2}$&$p_{3/2}^{-1/2}$&$p_{3/2}^{+1/2}$&$p_{3/2}^{-1/2}$&$p_{3/2}^{+3/2}$&$p_{3/2}^{-3/2}$&$p_{1/2}^{+1/2}$& & $\frac{3}{2}^{-}$ \\
$~^{11}$Be $(\frac{3}{2}^{+})$&$p_{3/2}^{+1/2}$&$p_{3/2}^{+3/2}$&$p_{3/2}^{+1/2}$&$p_{3/2}^{-1/2}$&$p_{3/2}^{+3/2}$&$p_{3/2}^{-3/2}$&$s_{1/2}^{-1/2}$& & $\frac{3}{2}^{+}$ \\
$~^{11}$Be $\frac{5}{2}^{-}$  &$p_{3/2}^{+1/2}$&$p_{3/2}^{+3/2}$&$p_{3/2}^{+1/2}$&$p_{3/2}^{-1/2}$&$p_{3/2}^{+3/2}$&$p_{3/2}^{-3/2}$&$p_{1/2}^{+1/2}$& & $\frac{5}{2}^{-}$ \\
    \hline
$~^{12}$Be,$0_{1}^{+}$        &$p_{3/2}^{+1/2}$&$p_{3/2}^{-1/2}$&$p_{3/2}^{+1/2}$&$p_{3/2}^{-1/2}$&$p_{3/2}^{+3/2}$&$p_{3/2}^{-3/2}$&$p_{1/2}^{+1/2}$&$p_{1/2}^{-1/2}$& $0^{+}$ \\
$~^{12}$Be,$2_{1}^{+}$        &$p_{3/2}^{+1/2}$&$p_{3/2}^{+3/2}$&$p_{3/2}^{+1/2}$&$p_{3/2}^{-1/2}$&$p_{3/2}^{+3/2}$&$p_{3/2}^{-3/2}$&$p_{1/2}^{+1/2}$&$p_{1/2}^{-1/2}$& $2^{+}$ \\
$~^{12}$Be,$0_{2}^{+}$        &$p_{3/2}^{+1/2}$&$p_{3/2}^{-1/2}$&$p_{3/2}^{+1/2}$&$p_{3/2}^{-1/2}$&$p_{3/2}^{+3/2}$&$p_{3/2}^{-3/2}$&$s_{1/2}^{+1/2}$&$s_{1/2}^{-1/2}$& $0^{+}$ \\
$~^{12}$Be,$1_{1}^{-}$        &$p_{3/2}^{+1/2}$&$p_{3/2}^{-1/2}$&$p_{3/2}^{+1/2}$&$p_{3/2}^{-1/2}$&$p_{3/2}^{+3/2}$&$p_{3/2}^{-3/2}$&$p_{1/2}^{+1/2}$&$s_{1/2}^{+1/2}$& $1^{-}$ \\
$~^{12}$Be,$(2_{1}^{-})$      &$p_{3/2}^{+1/2}$&$p_{3/2}^{-1/2}$&$p_{3/2}^{+1/2}$&$p_{3/2}^{-1/2}$&$p_{3/2}^{+3/2}$&$p_{3/2}^{-3/2}$&$p_{1/2}^{+1/2}$&$d_{5/2}^{+3/2}$& $2^{-}$ \\
$~^{12}$Be,$(2_{2}^{+})$      &$p_{3/2}^{+1/2}$&$p_{3/2}^{+3/2}$&$p_{3/2}^{+1/2}$&$p_{3/2}^{-1/2}$&$p_{3/2}^{+3/2}$&$p_{3/2}^{-3/2}$&$s_{1/2}^{+1/2}$&$s_{1/2}^{-1/2}$& $2^{+}$ \\
$~^{12}$Be,$(3_{1}^{-})$      &$p_{3/2}^{+1/2}$&$p_{3/2}^{+3/2}$&$p_{3/2}^{+1/2}$&$p_{3/2}^{-1/2}$&$p_{3/2}^{+3/2}$&$p_{3/2}^{-3/2}$&$p_{1/2}^{+1/2}$&$s_{1/2}^{+1/2}$& $3^{-}$ \\
    \hline
    \label{tab:iniwav}
\end{longtable}

\subsection{Table of numerical values}

\subsubsection{Energies}

In Table~\ref{tab:energy} we listed the numerical values of energies obtained by NLEFT
using N\(^3\)LO interaction~\cite{Elhatisari:2022zrb} and SU(4) interaction~\cite{Shen:2022bak}.

\begin{longtable}{|l|cc|c|}
    \caption{Energies of Be isotopes calculated by NLEFT using the SU(4) interaction \cite{Shen:2022bak}
    and N\(^3\)LO interaction \cite{Elhatisari:2022zrb},compared to
    experiment \cite{Liu:2022zcn,Wang:2021xhn,Kelley:2017qgh,Kelley:2012qua,Tilley:2004zz,Tilley:2002vg}.
    All energies are in MeV. The error bars of NLEFT results are one standard deviation estimates
    due to stochastic errors and Euclidean time extrapolation.} \\
    \hline
                                    &  SU(4) & N\(^3\)LO & Exp.      \\
    \hline
    $~^{7}$Be , $\frac{3}{2}^{-}$   & $-39.5(1)$ & $-38.5(1)$  & $-37.6$   \\
    $~^{7}$Be , $\frac{1}{2}^{-}$   & $-39.4(1)$ & $-37.8(4)$  & $-37.2$   \\
    $~^{7}$Be , $\frac{7}{2}^{-}$   & $-32.0(1)$ & $-32.0(1.2)$& $-33.0$   \\
    $~^{7}$Be , $(\frac{1}{2})^{+}$ & $-31.9(2)$ & $-28.8(1)$  & --   \\
    $~^{7}$Be , $(\frac{3}{2})^{+}$ & $-29.9(3)$ & $-30.5(8)$  & --   \\
    $~^{7}$Be , $(\frac{5}{2})^{+}$ & $-26.5(1)$ & $-26.5(7)$  & --   \\
    \hline
    $~^{8}$Be , $0_{1}^{+}$         & $-56.8(1)$ & $-56.7(4)$  & $-56.5$   \\
    $~^{8}$Be , $2_{1}^{+}$         & $-53.1(1)$ & $-53.6(7)$  & $-53.5$   \\
    $~^{8}$Be , $4_{1}^{+}$         & $-45.8(1.3)$& $-45.8(1.6)$& $-45.1$   \\
    $~^{8}$Be , $2_{1}^{-}$         & $-39.3(3)$ & $-39.7(4)$  & $-37.6$   \\
    \hline
    $~^{9}$Be , $\frac{3}{2}^{-}$   & $-56.1(1)$ & $-57.6(3)$  & $-58.2$   \\
    $~^{9}$Be , $\frac{1}{2}^{+}$   & $-56.4(1)$ & $-58.2(1)$  & $-56.5$   \\
    $~^{9}$Be , $\frac{5}{2}^{-}$   & $-54.0(1)$ & $-55.2(4)$  & $-55.7$   \\
    $~^{9}$Be , $\frac{1}{2}^{-}$   & $-55.9(2)$ & $-56.4(1)$  & $-55.4$   \\
    $~^{9}$Be , $\frac{5}{2}^{+}$   & $-52.9(1)$ & $-55.1(2)$  & $-55.1$   \\
    $~^{9}$Be , $\frac{3}{2}^{+}$   & $-51.4(4)$ & $-53.1(7)$  & $-53.5$   \\
    $~^{9}$Be , $\frac{7}{2}^{-}$   & $-49.1(1)$ & $-52.6(8)$  & $-51.8$   \\
    \hline
    $~^{10}$Be, $0_{1}^{+}$         & $-64.0(1)$ & $-63.7(3)$  & $-65.0$   \\
    $~^{10}$Be, $2_{1}^{+}$         & $-60.2(1)$ & $-62.6(1.2)$& $-61.6$   \\
    $~^{10}$Be, $2_{2}^{+}$         & $-59.1(1)$ & $-61.4(3)$  & $-59.0$   \\
    $~^{10}$Be, $1_{1}^{-}$         & $-56.7(2)$ & $-58.0(6)$  & $-59.0$   \\
    $~^{10}$Be, $0_{2}^{+}$         & $-57.6(2)$ & $-60.5(1.0)$& $-58.8$   \\
    $~^{10}$Be, $2_{1}^{-}$         & $-56.1(1)$ & $-59.5(3)$  & $-58.7$   \\
    $~^{10}$Be, $3_{1}^{-}$         & $-54.6(3)$ & $-57.6(8)$  & $-57.6$   \\
    $~^{10}$Be, $2_{3}^{+}$         & $-57.0(1)$ & $-59.9(8)$  & $-57.4$   \\
    $~^{10}$Be, $A_1^+(3)$          & $-56.1(7)$ & $-58.4(9)$  & --        \\
    $~^{10}$Be, $(4_{1}^{-})$       & $-52.6(1)$ & $-54.4(3)$  & $-55.7$   \\
    \hline
    $~^{11}$Be ,$\frac{1}{2}^{+}$   & $-64.6(1)$ & $-65.6(3)$  & $-65.5$   \\
    $~^{11}$Be ,$\frac{1}{2}^{-}$   & $-64.1(1)$ & $-63.9(1)$  & $-65.2$   \\
    $~^{11}$Be ,$\frac{5}{2}^{+}$   & $-60.5(2)$ & $-61.9(3)$  & $-63.7$   \\
    $~^{11}$Be ,$\frac{3}{2}^{-}$   & $-62.3(2)$ & $-62.7(1.0)$& $-62.8$   \\
    $~^{11}$Be ,$(\frac{3}{2}^{+})$ & $-59.9(1)$ & $-61.6(2)$  & $-62.1$\footnotemark[1]   \\
    $~^{11}$Be ,$\frac{5}{2}^{-}$   & $-60.5(3)$ & $-59.3(8)$  & $-61.6$   \\
    \hline
    $~^{12}$Be, $0_{1}^{+}$         & $-72.1(1)$ & $-67.9(4)$  & $-68.6$   \\
    $~^{12}$Be, $2_{1}^{+}$         & $-67.9(3)$ & $-65.9(1.6)$& $-66.5$   \\
    $~^{12}$Be, $0_{2}^{+}$         & $-66.8(2)$ & $-65.7(4)$  & $-66.4$   \\
    $~^{12}$Be, $1_{1}^{-}$         & $-65.2(1)$ & $-66.6(1)$  & $-65.9$   \\
    $~^{12}$Be, $(2_{1}^{-})$       & $-61.9(4)$ & $-61.8(9)$  & $-64.2$   \\
    $~^{12}$Be, $(2_{2}^{+})$       & $-64.1(3)$ & $-62.5(8)$  & $-63.8$   \\
    $~^{12}$Be, $(3_{1}^{-})$       & $-60.9(1)$ & $-61.7(1.4)$& $-62.9$\footnotemark[2]   \\
    \hline
    \label{tab:energy}
    \footnotetext[1]{For Exp., $(\frac{3}{2}^{+},\frac{3}{2}^{-})$}
    \footnotetext[2]{For Exp., $(4^{+}, 2^{+}, 3^{-})$}
\end{longtable}

\subsubsection{Radii}

In Table~\ref{tab:radius}, we give the values of the calculated charge and matter radii in
comparison to experiment.

\begin{table}[!htp]
    \centering
    \caption{Energies of Be isotopes calculated by NLEFT using the SU(4) interaction \cite{Shen:2022bak}
    and N\(^3\)LO interaction \cite{Elhatisari:2022zrb},compared to experiment.
    All energies are in MeV and radii in fm.
      For the NLEFT results, the error bars are one standard deviation estimates due to stochastic errors and
Euclidean time extrapolation.}
    \begin{tabular}{|l|cc|c|}
    \hline
    $r_c$ (fm) & SU(4) & N\(^3\)LO &  Exp. \cite{Krieger:2012jx,Nortershauser:2008vp}  \\
    \hline                  
    $~^{7}$Be  & 2.557(3)   & 2.580(18)   & 2.647(17) \\
    $~^{9}$Be  & 2.576(8)   & 2.552(16)   & 2.519(12) \\
    $~^{10}$Be & 2.466(5)   & 2.511(37)   & 2.357(18) \\
    $~^{11}$Be & 2.576(4)   & 2.543(41)   & 2.463(16) \\
    $~^{12}$Be & 2.352(2)   & 2.579(34)   & 2.503(15) \\
    \hline                  
    $r_m$ (fm) & NLEFT, SU(4) & NLEFT, N\(^3\)LO & Exp. \cite{Dobrovolsky:2019wzt,Tanihata:2013jwa} \\
    \hline                  
    $~^{7}$Be  & 2.37(1)    & 2.39(1)     & 2.42(4)  \\
    $~^{9}$Be  & 2.60(1)    & 2.52(1)     & 2.38(1)  \\
    $~^{10}$Be & 2.48(1)    & 2.53(2)     & 2.30(2)  \\
    $~^{11}$Be & 3.14(1)    & 2.86(1)     & 2.91(5)  \\
    $~^{12}$Be & 2.45(1)    & 2.63(1)     & 2.59(6)  \\
    \hline
    \end{tabular}
    \label{tab:radius}
\end{table}

\subsubsection{Comparison of transition properties}

In Table~\ref{tab:s-trans}, we give the values of the calculated transition properties in
comparison to experiment.

\begin{table}[!h]
\scalebox{0.8}{
\begin{tabular}{lc|cccccccc}
\hline
$~^{ 7}$Be    & Exp. \cite{Henderson:2019ubp}  & N\(^3\)LO & SU(4) & NCSMC \cite{Vorabbi:2019imi} & NCFC \cite{Heng:2016umo} & GFMC \cite{Pervin:2007sc,Pastore:2012rp} &  \\
\hline
$E2,\frac{3}{2}^{-}\to \frac{1}{2}^{-}$ & 26(6)(3) & 15.2(5) & 16.0(2) & 20.0 & 19.3 & 22.2(11)$-$27.5(8) &        \\
\hline
$~^{ 9}$Be                       & Exp. \cite{Tilley:2004zz}  & NLEFT & SU(4) & CSM \cite{DellaRocca:2018mrt} & AMD \cite{Kanada-Enyo:2015knx} & GFMC \cite{Pastore:2012rp,Pieper:2002ne} & NCSM \cite{Forssen:2004dk} & MMM \cite{Arai:2003jm}    \\
\hline
$Q(\frac{3}{2}^{-})$                    & 5.29(4) \cite{Stone:2016bmk}  & 7.4(1.0)  & 7.3(1)    & 5.30  &       & 5.0(3)$-$8.5(3)  & &        \\
$E1,\frac{1}{2}^{+}\to \frac{3}{2}^{-}$ & 0.136(2) \cite{Arnold:2011nv} & 0.060(15) & 0.131(3)  & 0.048 & 0.002 &         & 0.033 & 0.061   \\
$E1,\frac{5}{2}^{+}\to \frac{3}{2}^{-}$ & 0.010(8)                      & 0.049(5)  & 0.045(14) & 0.005 & 0.013 &         & 0.006 & 0.025   \\
$E2,\frac{5}{2}^{-}\to \frac{3}{2}^{-}$ & 27.1(2.0)                     & 27.8(1.9) & 35.7(1.8) & 23.9  &       & 25.6(6) &       & 17.6(6) \\
$E2,\frac{7}{2}^{-}\to \frac{3}{2}^{-}$ & 9.5(4.1)                      & 5.3(8)    & 11.6(2.5) & 10.0  &       &         &       &  \\
\hline
$~^{10}$Be                       & Exp. \cite{Tilley:2004zz}  & NLEFT & SU(4) & Multicool \cite{Myo:2023alz} & GCM \cite{Descouvemont:2020kwq} & NCSM \cite{Orce:2012yc} & MCSM \cite{Liu:2011xv} & GFMC \cite{McCutchan:2009th} &  MO \cite{Itagaki:1999vm}       \\
\hline
$E1,3_1^{-}\to 2_1^{+}$& 0.009(1)                       & 0.004(3)& 0.026(2) &    &    &   &    &                &     \\
$E2,2_1^{+}\to 0_1^{+}$&  9.2(3) \cite{McCutchan:2009th}&8.5(9) & 10.6(4)  &7.9 &5.66&9.8&9.3 &8.1(3)$-$17.9(5)&11.26\\
\hline
$~^{11}$Be                       & Exp. \cite{Kwan:2014dha}  & NLEFT & SU(4) &  NCSMC \cite{Calci:2016dfb} & AMD \cite{Dan:2021htj} & GCM \cite{Descouvemont:2020kwq}  \\
\hline
$E1,\frac{1}{2}^{-}\to \frac{1}{2}^{+}$ & 0.102(2) & 0.038(3) & 0.023(3) & 0.117-0.146 & 0.61-0.73 & 0.002 \\
\hline
$~^{12}$Be                    & Exp.  & NLEFT & SU(4) & FMD \cite{Krieger:2012jx} & GCM \cite{Dufour:2010dmf} & NCSM \cite{Dufour:2010dmf} & HAEM \cite{Romero-Redondo:2008hux} & AMD \cite{Kanada-Enyo:2003fhn} \\
\hline
$E1,0_{1}^{+}\to 1_{1}^{-}$&0.051(13)\cite{Iwasaki:2000gp}    &0.056(26)&0.049(2)&    &    &         & 0.046$-$0.064&   \\
$E2,2_{1}^{+}\to 0_{1}^{+}$&14.2(1.0)(2.0)\cite{Morse:2018ojw}&9.0(3.1) &7.8(1.1)&8.75&12.6&3.5$-$4.6& 3.04$-$3.93  &14 \\
\hline
\end{tabular}
}
\caption{
Quadrupole moments and transition rates of Be isotopes calculated using NLEFT with the N\(^3\)LO interaction~\cite{Elhatisari:2022zrb} and the SU(4) interaction~\cite{Shen:2022bak}, compared to experimental data and other theoretical models. Units: $Q$ and $m(E0)$ in $e$fm$^2$, $B(E1)$ in $e^2$fm$^2$, and $B(E2)$ in $e^2$fm$^4$. For results presented as ranges, the ranges reflect variations from different interactions and numerical methods. \label{tab:s-trans}
}
\end{table}

\subsection{Intrinsic density}

There is no fixed definition of the one-body intrinsic density, $\rho(\mathbf{r})$, derived from the many-body density, $\rho(\mathbf{r}_1, \mathbf{r}_2, \dots, \mathbf{r}_A)$. A common approach is to align each configuration along the principal axis \cite{Wiringa:2000gb}. However, this alignment can artificially enhance the prominence of the principal axis. In our study of neutron-rich Beryllium isotopes, we observe that this method tends to position the valence neutrons along the long principal axis, as they are often located far from the center.

Building on our previous work with $^{12}$C \cite{Shen:2022bak}, we introduce an \textit{$\alpha$-cluster-view intrinsic density}, illustrated in schematic Figure~\ref{fig:sch}. Similar to Ref.~\cite{Shen:2022bak}, we first group the nucleons into $N_\alpha$ clusters by considering all possible permutations of protons and neutrons and selecting the arrangement that minimizes the sum of inter-nucleon distances. We then determine the symmetry axis for each configuration; however, this axis is not uniquely defined and depends on the specific system under investigation.
For example, in $^{8}$Be, identifying the symmetry axis as the $z$-axis is straightforward because the two clusters always align linearly. In contrast, for $^{9}$Be, the presence of a valence neutron introduces an angle between the clusters, resulting in a angle $(\alpha_1, 0, \alpha_2) < 180^\circ$. This approach can be generalized to other nuclei, such as the ground state of $^{12}$C, where the symmetry axis is defined based on an equilateral triangular arrangement of three clusters \cite{Shen:2022bak}. However, in each configuration, two clusters are typically closer together than the third, allowing the symmetry axis to also be defined based on an acute triangular arrangement rather than an equilateral one. Therefore, the choice of symmetry axis depends on the specific problem and the perspective adopted.
In this study, we consistently define the symmetry axis of Beryllium isotopes as the $z$-axis and randomly select one cluster to align along the positive or negative $z$ direction.

\begin{figure}[!htbp]
  \centering
  \includegraphics[width=0.49\textwidth]{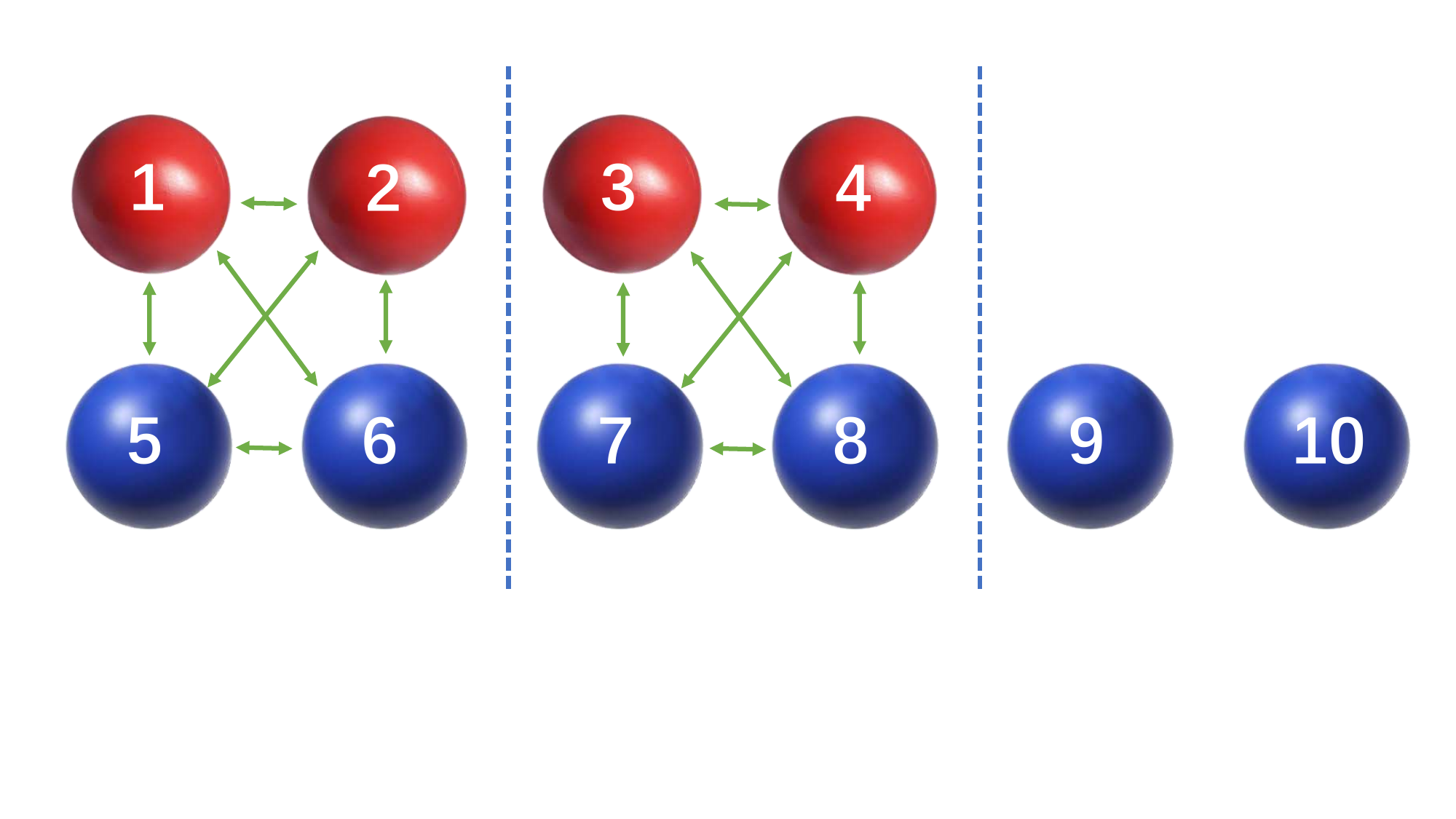}
  \includegraphics[width=0.49\textwidth]{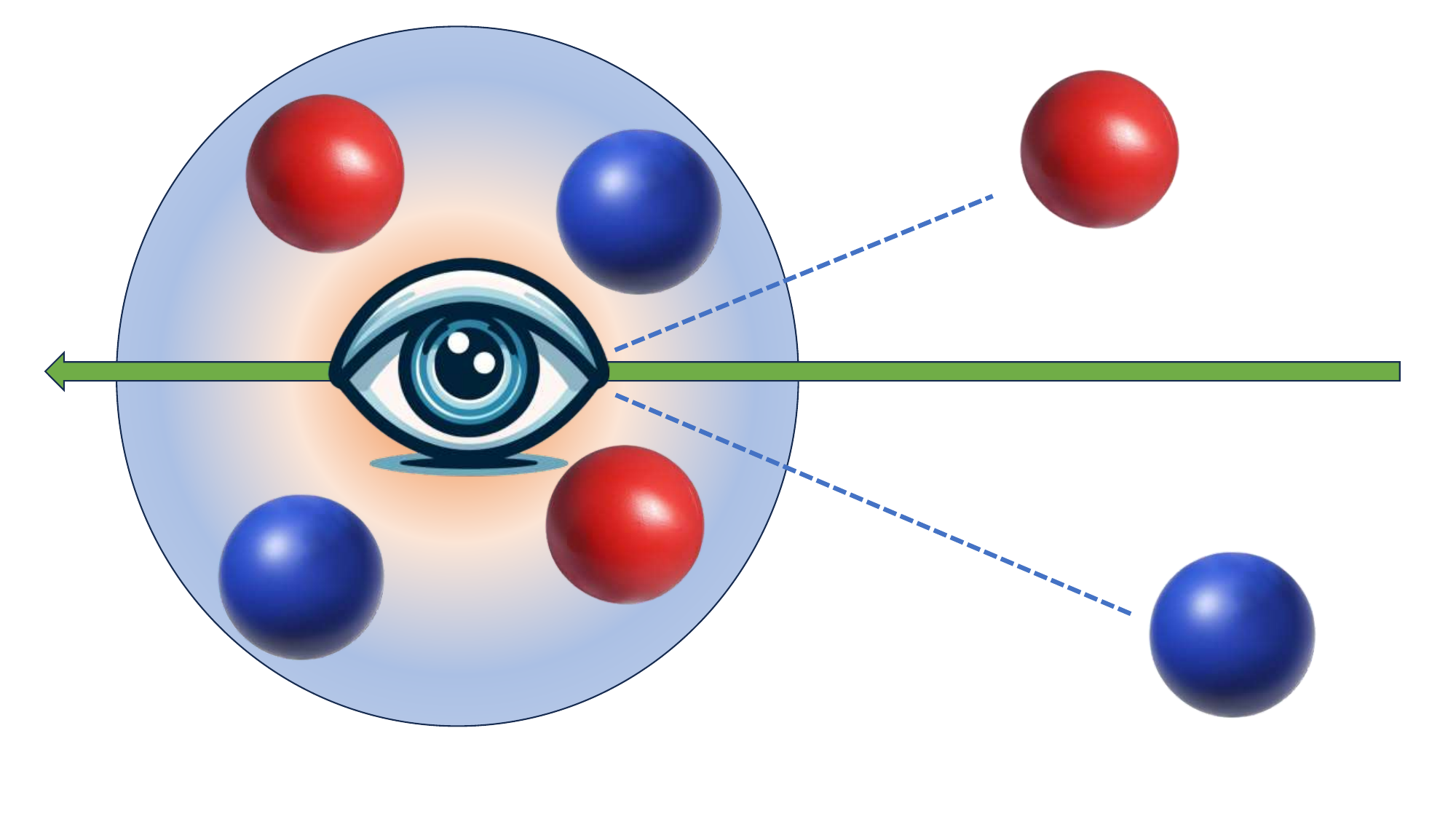}
  \caption{Schematic show of grouping clusters and the intrinsic density from the view of cluster.}
  \label{fig:sch}
\end{figure}

In Figure~\ref{fig:tot}, we present the total density derived using the previously introduced \textit{$\alpha$-cluster-view intrinsic density}. The general features observed from $^{8}$Be to $^{12}$Be have been discussed in the main text and show good agreement with results from other studies.
Specifically, for $^{7}$Be, the characteristic \(^4\)He plus \(^3\)He clustering is effectively captured by the \textit{$\alpha$-cluster-view intrinsic density}. In the case of the ambiguous positive-parity state of $^{7}$Be, one proton is excited out of the \(^3\)He cluster. This excitation disrupts the small cluster, leading to significant shape deformation and spatial extension, as illustrated by the $\beta_{\rm pin}$ and $\gamma_{\rm pin}$ plots in the main text.
Similarly, for the \(2_1^-\) state of $^{8}$Be, one proton is excited out of the \(\alpha\) cluster. This results in a very high excitation energy and a substantial change in the nuclear shape.
In the present study, the third component of the total spin is fixed at $J_z = J$. Exploring the $J_z$-dependence of the structure would be an interesting avenue for future investigation.

\begin{figure}[!htbp]
  \centering
  \includegraphics[width=\textwidth]{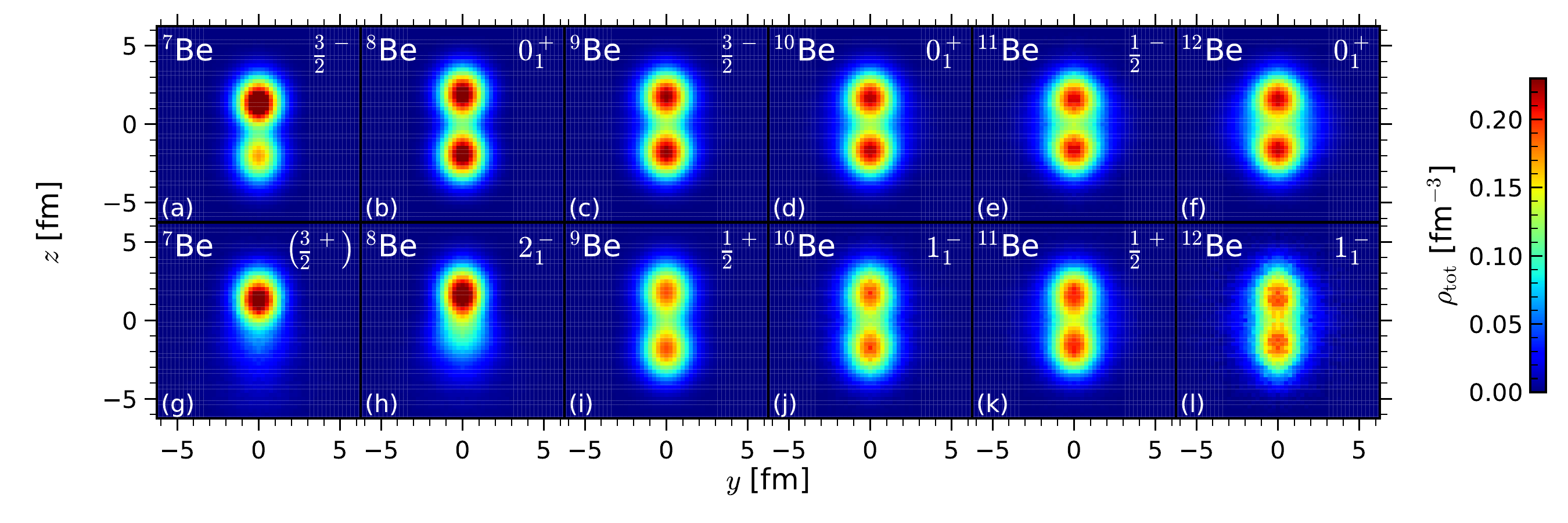}
  \caption{Intrinsic density at $x = 0$ plane of selected states of beryllium isotopes obtained by NLEFT using N\(^3\)LO interaction. The third component of total spin is fixed at $J_z = J$.}
  \label{fig:tot}
\end{figure}

In Figure~\ref{fig:val}, we display the density distributions of valence particles obtained using the previously introduced \textit{$\alpha$-cluster-view intrinsic density}. For $^{7}$Be, these valence particles consist of two protons and one neutron that are not grouped into the $\alpha$ cluster. In contrast, for the other isotopes, the valence particles are neutrons. To maintain a consistent color scale across all panels, we normalized the density distributions differently for each case. For example, in the $^{7}$Be $3/2^-$ state, the $^{3}$He cluster exhibits a much higher concentration, necessitating division by a larger normalization factor.
The features of the $\pi$ and $\sigma$ orbitals, which are discussed in detail in the main text, are more clearly observable in the $^{9}$Be case, as shown in panels (b) and (d). The low probability regions at small $z$ and large $y$ coordinates may be attributed to the finite Euclidean time projection used in our calculations. Due to the sign problem, the computations cannot extend to larger $L_t$ values and are limited to $L_t = 300$.
Remarkably, starting from a common shell-model initial wavefunction, distinct nuclear molecular orbitals emerge automatically. The severity of the sign problem generally increases with the number of nucleons and for states with unnatural parity, resulting in larger errors in the valence particle densities compared to the total densities.
Consequently, the uneven density distributions plotted, such as those in panel (j) of Fig.~\ref{fig:val}, are purely due to statistical errors.

\begin{figure}[!htbp]
  \centering
  \includegraphics[width=\textwidth]{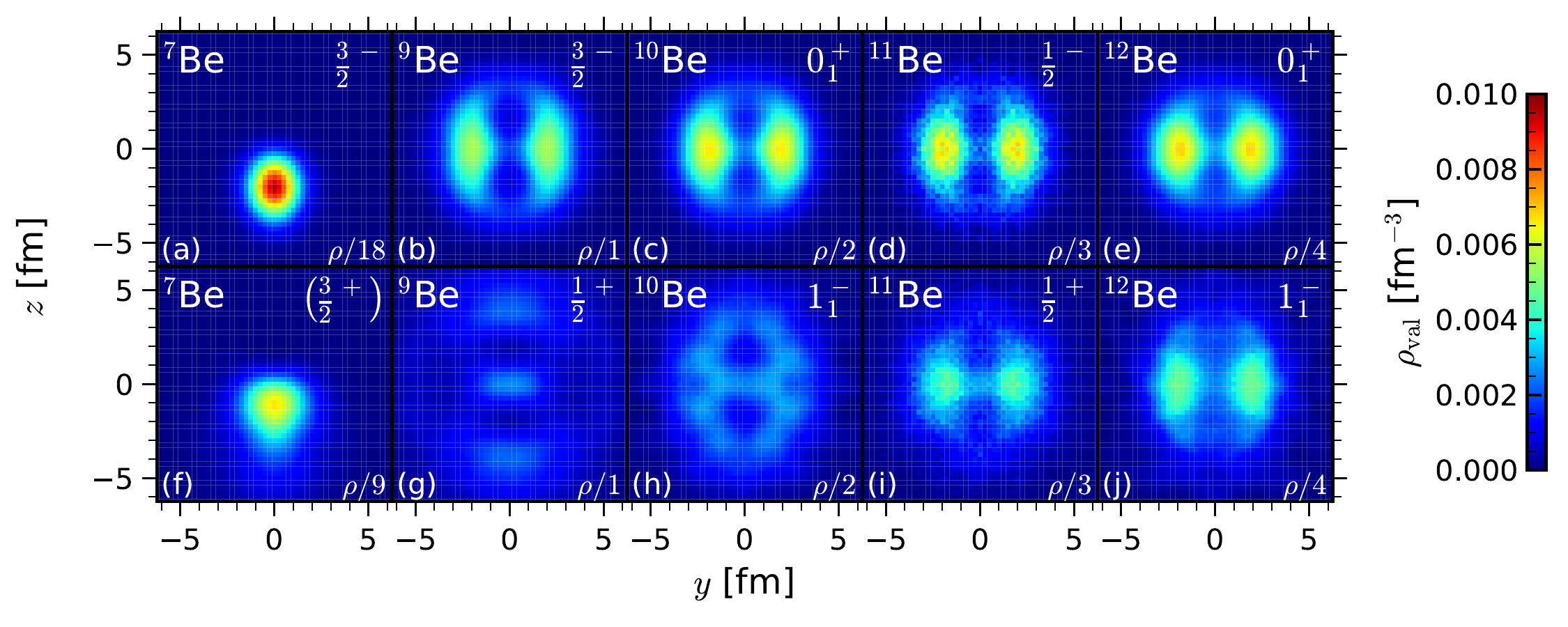}
  \caption{Intrinsic density at $x = 0$ plane of selected states of beryllium isotopes obtained by NLEFT using N\(^3\)LO interaction. The third component of total spin is fixed at $J_z = J$.}
  \label{fig:val}
\end{figure}

\subsection{Molecular orbitals}

In atomic and molecular physics, the molecular orbital describes how electrons moves around the composite atoms of a molecule.
Due to the strong binding character, the $\alpha$ particle behaves like an atom and several $\alpha$ particles can form a similar
structure of a 'nuclear molecule' with extra nucleons move around in 'molecular orbitals'.
The Beryllium isotopes is a classical example of showing this picture \cite{von1996two}.
While the two $\alpha$ particles do not form a bound nucleus of $^{8}$Be, adding an extra neutron
in a $\pi$ orbital inbetween brings the two $\alpha$ particles together and forms a bound nucleus of $^{9}$Be.

Here we would like to take a simple toy model to show the basic features of those molecular orbitals, in comparison
with our \textit{ab initio} findings.
We assume two $\alpha$ particles located at $z = \pm 1.6$ fm with 4 Gaussian wave packets each.
The following single-particle orbitals are then added
\begin{subequations}\label{eq:}\begin{align}
  \psi_\pi &= e^{-c_1(\mathbf{r}-\mathbf{r}_z)^2} (x+iy) + e^{-c_1(\mathbf{r}+\mathbf{r}_z)^2} (x+iy), \\
  \psi_\sigma &= c_2 e^{-c_3\mathbf{r}^2} + c_4 e^{-c_1(\mathbf{r}-\mathbf{r}_z)^2} (z-|\mathbf{r}_z|)
  - c_4 e^{-c_1(\mathbf{r}+\mathbf{r}_z)^2} (z+|\mathbf{r}_z|),
\end{align}\end{subequations}
with $\mathbf{r}_z$, $c_1$ to $c_4$ are parameters to adjust to better agree with \textit{ab initio} results.
Using pinhole algorithm we can obtain the densities of the system made up of these wave functions
with proper orthonormalization condition.
From panel (a) to (e) in Fig.~\ref{fig:molecular}, orbitals of $1\pi, 1\sigma, 2\pi, 2\pi+1\sigma$, and $2\pi+2\sigma$ are
added respectively.
The total density are plotted to be compared to
$^{9}$Be $\frac{3}{2}^-$, 
$^{9}$Be $\frac{1}{2}^+$, 
$^{10}$Be $0^+$, 
$^{11}$Be $\frac{1}{2}^+$, and
$^{12}$Be $0^+$ in Fig.~\ref{fig:tot}.
From panel (f) to (j) in Fig.~\ref{fig:molecular}, the densities of corresponding valence orbitals normalized to 1 are
plotted to be compared to those in Fig.~\ref{fig:val}.

\begin{figure}[!htbp]
  \centering
  \includegraphics[width=0.99\textwidth]{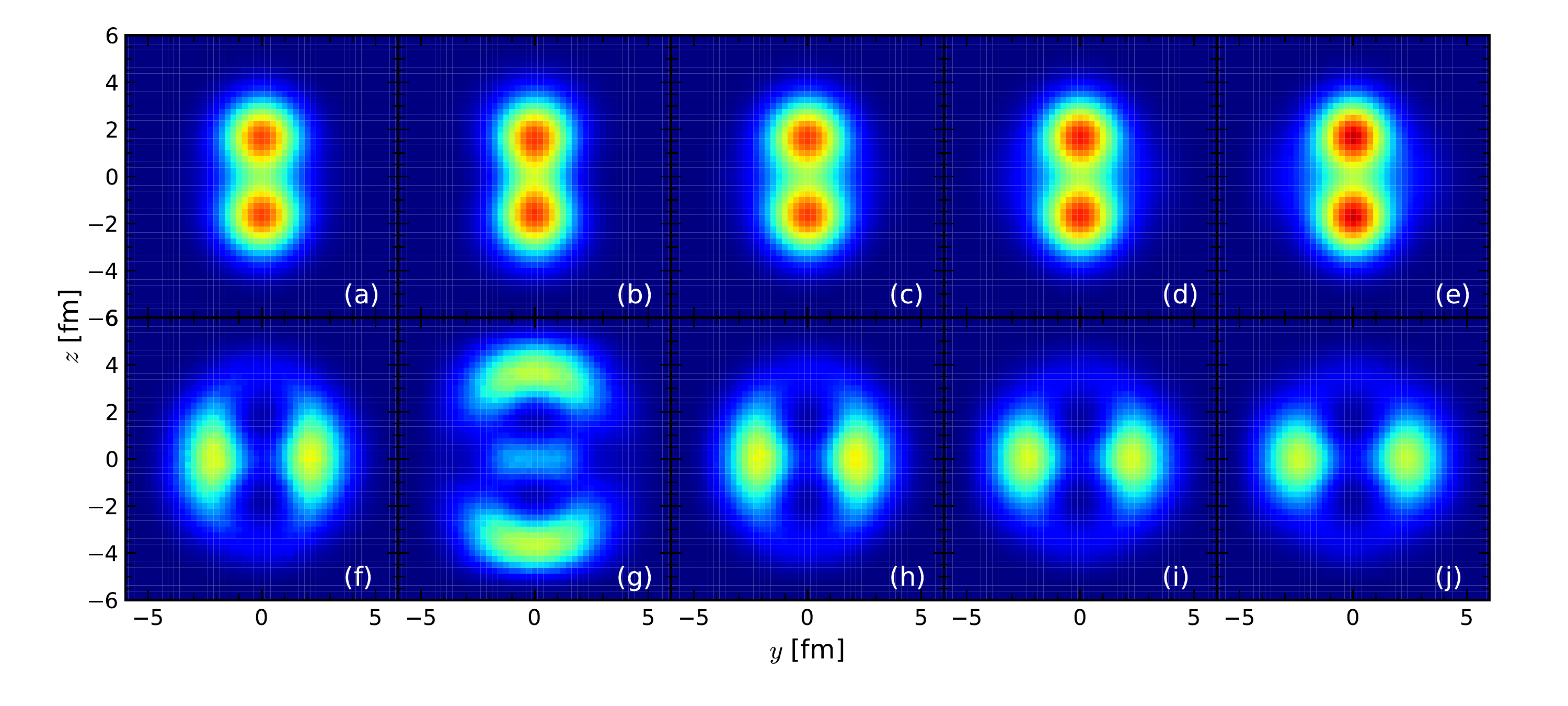}
  \caption{Schematic show of density plot of $\pi$ and $\sigma$ orbitals by adding single-particle orbitals around
  two $\alpha$ clusters. Panel (a) to (e) are total densities, and (f) to (j) are densities of corresponding valence orbitals.
  See text for detail description.}
  \label{fig:molecular}
\end{figure}

With the simplified setup we can already observe many basic features of the \textit{ab initio} results
in Fig.~\ref{fig:tot} and \ref{fig:val} such as the two-center clustering,
$\pi$-orbital, and $\sigma$-orbital.
For more detailed information of nuclear moleculars the readers are refered to Ref.~\cite{von2006nuclear}.
While the molecular orbitals can be separated easily in such simple toy models, it is not trivial to identify them
given a full many-body correlated wave function $\Psi(\mathbf{r}_1,\mathbf{r}_2, \dots ,\mathbf{r}_A)$.
The combination of pinhole algorithm and the identification
of $\alpha$ clusters and valence particles provides a powerful way to achieve this goal.
In this way there is no need to assume clusters and valence particles in the beginning, they
emerged naturally from the full wave function $\Psi(\mathbf{r}_1,\mathbf{r}_2, \dots ,\mathbf{r}_A)$.
The \textit{ab initio} density plots can be used in this manner to deduce the nature
and details of the molecular orbitals in nuclei with clustering.

\end{onecolumngrid}
\end{appendix}


\putbib[bref-sm]

\end{bibunit}

\end{document}